\pdfoutput=1
\documentclass[final,0p,times]{elsarticle}

\usepackage{graphicx}
\usepackage{multirow}
\usepackage{tocloft}

\usepackage{bm}
\usepackage{amsmath}
\usepackage{amssymb}
\usepackage{amsthm}

\newcommand{\IGNORE}[1]{}
\newcommand{\ignore}[1]{}
\newcommand{\veps}{\varepsilon}

\newcommand{\opn}{\operatorname}
\newcommand{\re}{\operatorname{Re}}
\newcommand{\im}{\operatorname{Im}}

\newcommand{\phm}{\phantom{-}}

\newcommand{\mc}[1]{\mathcal{#1}}

\newcommand{\jd}{\displaystyle}
\newcommand{\js}{\scriptstyle}
\newcommand{\jt}{\textstyle}

\newcommand{\der}[2]{\frac{\partial #1}{\partial #2}}
\newcommand{\derr}[2]{\frac{\delta #1}{\delta #2}}

\newcommand{\wtil}[1]{\widetilde{#1}}

\newcommand{\wdg}[1]{{#1}^{\scriptscriptstyle\bm\wedge}}

\newcommand{\x}{\tilde{x}}

\newcommand{\e}[1]{{(#1)}}
\newcommand{\hh}{h}

\makeatletter
\def\imod#1{\allowbreak\mkern10mu({\operator@font mod}\,\,#1)}
\makeatother

\journal{Computational Science and Discovery}

\begin{document}

\begin{frontmatter}

\title{Overdetermined Shooting Methods for Computing \\
  Standing Water Waves with Spectral Accuracy}

\author[berk]{Jon Wilkening}
\author[berk]{Jia Yu}

\address[berk]{Department of Mathematics, University of California,
Berkeley}

\begin{abstract}

  A high-performance shooting algorithm is developed to compute
  time-periodic solutions of the free-surface Euler equations with
  spectral accuracy in double and quadruple precision. The method is
  used to study resonance and its effect on standing water waves.
  We identify new nucleation mechanisms in which isolated
  large-amplitude solutions, and closed loops of such solutions,
  suddenly exist for depths below a critical threshold.  We also study
  degenerate and secondary bifurcations related to Wilton's ripples in
  the traveling case, and explore the breakdown of self-similarity at
  the crests of extreme standing waves.  In shallow water, we find
  that standing waves take the form of counter-propagating solitary
  waves that repeatedly collide quasi-elastically. In deep water with
  surface tension, we find that standing waves resemble
  counter-propagating depression waves.  We also discuss existence and
  non-uniqueness of solutions, and smooth versus erratic dependence
  of Fourier modes on wave amplitude and fluid depth.
  
  In the numerical method, robustness is achieved by posing the
  problem as an overdetermined nonlinear system and using either
  adjoint-based minimization techniques or a quadratically convergent
  trust-region method to minimize the objective function.  Efficiency
  is achieved in the trust-region approach by parallelizing the
  Jacobian computation so the setup cost of computing the
  Dirichlet-to-Neumann operator in the variational equation is not
  repeated for each column.  Updates of the Jacobian are also delayed
  until the previous Jacobian ceases to be useful.  Accuracy is
  maintained using spectral collocation with optional mesh refinement
  in space, a high order Runge-Kutta or spectral deferred correction
  method in time, and quadruple-precision for improved navigation of
  delicate regions of parameter space as well as validation of
  double-precision results.  Implementation issues for GPU
  acceleration are briefly discussed, and the performance of the
  algorithm is tested for a number of hardware configurations.

\end{abstract}


\end{frontmatter}

% \linenumbers

\vspace*{-18pt}
\tableofcontents

\thispagestyle{empty}

\section{Introduction}

Time-periodic solutions of the free-surface Euler equations serve as
an excellent benchmark for the design and implementation of numerical
algorithms for two-point boundary value problems governed by nonlinear
partial differential equations.  In particular, there is a large body
of existing work on numerical methods for computing standing waves
\cite{schwartz:81, mercer:92, mercer:94, smith:roberts:99,
  tsai:jeng:94, bryant:stiassnie:94, schultz, okamura:03} and
short-crested waves \cite{roberts83b, Marchant87b, bridges01,
  Ioualalen06} for performance comparison.
Moreover, many of these previous studies reach contradictory
scientific conclusions that warrant further investigation, especially
concerning extreme waves and the formation of a corner or cusp.
Penney and Price \cite{penney:52} predicted a 90 degree corner, which
was verified experimentally by G. I. Taylor \cite{taylor:53}, who was
nevertheless skeptical of their analysis.  Grant \cite{grant} and
Okamura \cite{okamura:98} gave theoretical arguments supporting the 90
degree corner.  Schwartz and Whitney \cite{schwartz:81} and Okamura
\cite{okamura:03} performed numerical experiments that backed the 90
degree conjecture.  Mercer and Roberts \cite{mercer:92} predicted a
somewhat sharper angle and mentioned 60 degrees as a possibility.
Schultz \emph{et al.}~\cite{schultz} obtained results similar to
Mercer and Roberts, and proposed that a cusp may actually form rather
than a corner.  Wilkening \cite{breakdown} showed that extreme
waves do not approach a limiting wave at all due to fine scale
structure that emerges at the surface of very large amplitude waves
and prevents the wave crest from sharpening in a self-similar manner.
This raises many new questions about the behavior of large-amplitude
standing waves, which we will explore in Section~\ref{sec:breakdown}.

% These discrepancies appear to be caused by loss of accuracy due to
% insufficient numerical resolution in previous studies.

On the theoretical side, it has long been known
\cite{concus:64,vanden:broeck:book,iooss05} that standing water waves
suffer from a small-divisor problem that obstructs convergence of the
perturbation expansions developed by Rayleigh \cite{rayleigh1876},
Penney and Price \cite{penney:52}, Tadjbakhsh and Keller
\cite{tadjbakhsh}, Concus \cite{concus:62}, Schwartz and Whitney
\cite{schwartz:81}, and others.  Penney and Price \cite{penney:52}
went so far as to state, ``there seems little likelihood that a proof
of the existence of the stationary waves will ever be given.''
Remarkably, Plotnikov and Toland \cite{plotnikov01}, together with
Iooss \cite{iooss05}, have recently established existence of
small-amplitude standing waves using a Nash-Moser iteration. As often
happens in small-divisor problems \cite{craig:wayne,bourgain99},
solutions could only be proved to exist for values of an amplitude
parameter in a totally disconnected Cantor set. No assertion is made
about parameter values outside of this set. This raises intriguing new
questions about whether resonance really causes a complete loss of
smoothness in the dependence of solutions on amplitude, or if these
results are an artifact of the use of Nash-Moser theory to prove
existence.  While a complete answer can only come through further
analysis, insight can be gained by studying high precision numerical
solutions.

In previous numerical studies, the most effective methods for
computing standing water waves have been Fourier collocation in space
and time \cite{vandenBroeck:81, tsai:jeng:94, okamura:03,
  ioualalen:03, okamura:10}, semi-analytic series expansions
\cite{schwartz:81,amick:87}, and shooting methods \cite{mercer:92,
  mercer:94, schultz, smith:roberts:99}.  In Fourier collocation,
time-periodicity is built into the basis, and the equations of motion
are imposed at collocation points to obtain a large nonlinear system
of equations.  This is the usual approach taken in analysis to prove
existence of time-periodic solutions, e.g.~of nonlinear wave equations
\cite{craig:wayne} or nonlinear Schr\"odinger equations
\cite{bourgain99}. The drawback as a numerical method is that the
number of unknowns in the nonlinear system grows like $(\Delta
x\,\Delta t)^{-1}$ rather than $\Delta x^{-1}$ for a shooting method,
which limits the resolution one can achieve.  Orthogonal collocation,
as implemented in the software package AUTO \cite{doedel91}, would be
less efficient than Fourier collocation as more timesteps will be
required to achieve the same accuracy.

The semi-analytic series expansions of Schwartz and Whitney
\cite{schwartz:81, amick:87} are a significant improvement over
previous perturbation methods \cite{rayleigh1876, penney:52,
  tadjbakhsh, concus:62} in that the authors show how to compute an
arbitrary number of terms rather than stopping at 3rd or 5th order.
They also used conformal mapping to flatten the boundary, which leads
to a more promising representation of the solution of Laplace's
equation.  As a numerical method, the coefficients of the expansion
are expensive to compute, which limits the number of terms one can
obtain in practice. (Schwartz and Whitney stopped at 25th order). It
may also be that the resulting series is an asymptotic series rather
than a convergent series.  Nevertheless, these series expansions play
an essential role in the proof of existence of standing waves on deep
water by Plotnikov, Toland and Iooss \cite{iooss05}.

In a shooting method, one augments the known boundary values at one
endpoint with additional prescribed data to make the initial value
problem well posed, then looks for values of the new data to satisfy
the boundary conditions at the other endpoint.  For ordinary
differential equations, this normally leads to a system of equations
with the same number of equations as unknowns. The same is true of
multi-shooting methods \cite{keller:68, stoer:bulirsch,
  guckenheimer:00}.  When the boundary value problem is governed by a
system of partial differential equations, it is customary to
discretize the PDE to obtain an ODE, then proceed as described above.
However, because of aliasing errors, quadrature errors, filtering
errors, and amplification by the derivative operator, discretization
causes larger errors in high-frequency modes than low-frequency modes
when the solution is evolved in time. These errors can cause the
shooting method to be too aggressive in its search for initial
conditions, and to explore regions of parameter space (the space of
initial conditions) where either the numerical solution is
inaccurate, or the physical solution becomes singular before
reaching the other endpoint. Even if safeguards are put in place to
penalize high-frequency modes in the search for initial conditions,
the Jacobian is often poorly conditioned due to these discretization
errors.

We have found that posing boundary value problems governed by PDEs as
overdetermined, nonlinear least squares problems can dramatically
improve the robustness of shooting methods in two critical ways.
First, we improve accuracy by padding the initial condition with
high-frequency modes that are constrained to be zero. With enough padding,
all the degrees of freedom controlled by the shooting method can be
resolved sufficiently to compute a reliable Jacobian. Second, adding
more rows to the Jacobian increases its smallest singular values,
often improving the condition number by several orders of magnitude.
The extra rows come from including the high-frequency modes of the
boundary conditions in the system of equations, even though they are
not included in the list of augmented initial conditions. As a rule of
thumb, it is usually sufficient to set the top 1/3 to 1/2 of the
Fourier spectrum to zero initially; additional zero-padding has little
effect on the numerical solution or the condition number.  Validation
of accuracy by monitoring energy conservation and decay rates of
Fourier modes will be discussed in Section~\ref{sec:breakdown}, along
with mesh refinement studies and comparison with quadruple precision
calculations.

In this paper, we present two methods of solving the nonlinear least
squares problem that arises in the overdetermined shooting framework.
The first is the adjoint continuation method (ACM) of Ambrose and
Wilkening \cite{benj1,benj2,vtxs1,lasers}, in which the gradient of
the objective function with respect to initial conditions is computed
by solving an adjoint PDE, and the BFGS algorithm \cite{bfgs, nocedal}
is used for the minimization.  This was the approach used by one of
the authors in her dissertation \cite{jia:thesis} to obtain the
results of Sections~\ref{sec:unit} and~\ref{sec:surf}.
In the second approach, we exploit an opportunity for parallelism that
makes computing the entire Jacobian feasible. Once this is done, a
variant of the Levenberg-Marquardt method (with less frequent Jacobian
updates) is used to rapidly converge to the solution. The main
challenge here is organizing the computation to maximize re-use of
setup costs in solving the variational equation with multiple
right-hand sides, to minimize communication between threads or with
the GPU device, and to ensure that most of the linear algebra occurs
at level 3 BLAS speed.  The performance of the algorithms on various
platforms is reported in Section~\ref{sec:perform}.

The scientific focus of the present work is on resonance and its
effect on existence, non-uniqueness, and physical behavior of standing
water waves.  A summary of our main results is given in the abstract,
and in more detail at the beginning of Section~\ref{sec:results}.  We
mention here that resonant modes generally take the form of
higher-frequency, secondary standing waves oscillating at the surface
of larger-scale, primary standing waves.  Because the equations are
nonlinear, only certain combinations of amplitude and phase can occur
for each component wave.  This leads to non-uniqueness through
multiple branches of solutions.  In shallow water,
bifurcation curves of high-frequency Fourier modes behave
erratically and contain many gaps where solutions do not appear to
exist.  This is expected on theoretical grounds.  However, these
bifurcation ``curves'' become smoother, or ``heal,'' as fluid depth
increases.  In infinite depth, such resonant effects are largely
invisible, which we quantify and discuss in Section~\ref{sec:conclude}.

In future work \cite{water:stable}, the methods of this paper will be
used to study other families of time-periodic solutions of the
free-surface Euler equations with less symmetry than is assumed here,
e.g.~traveling-standing waves, unidirectional solitary wave
interactions, and collisions of gravity-capillary solitary waves. The
stability of these solutions will also be analyzed in
\cite{water:stable} using Floquet methods.

\section{Equations of motion and time-stepping}
\label{sec:eqs}

The effectiveness of a shooting algorithm for solving two-point
boundary value problems is limited by the accuracy of the
time-stepper.  In this section, we describe a boundary integral
formulation of the water wave problem that is spectrally accurate in
space and arbitrary order in time.  We also describe how to implement
the method in double and quadruple precision using a GPU, and discuss
symmetries of the problem that can be exploited to reduce the work of
computing standing waves by a factor of 4. The method is similar to
other boundary integral formulations \cite{lh76, baker:82, krasny:86,
  mercer:92, mercer:94, smith:roberts:99, baker10}, but is simpler to
implement than the angle--arclength formulation used in \cite{hls94,
  ceniceros:99, HLS01, vtxs1}, and avoids issues of identifying two
curves that are equal ``up to reparametrization'' when the $x$ and $y$
coordinates of the interface are both evolved (in non-symmetric
problems). Our approach also avoids sawtooth instabilities that
sometimes occur when using Lagrangian markers
\cite{lh76,mercer:92}. This is consistent with the results of Baker
and Nachbin \cite{baker:nachbin:98}, who found that sawtooth
instabilities can be controlled without filtering using the correct
combination of spectral differentiation and interpolation
schemes. While conformal mapping methods \cite{dyachenko:1996,
  nachbin:04, milewski:11} are more efficient than boundary integral
methods in many situations, they are not suitable for modeling extreme
waves as the spacing between grid points expands severely in regions
where wave crests form, which is the opposite of what is needed for an
efficient representation of the solution via mesh refinement.

\subsection{Equations of motion}\label{sec:motion}

We consider a two-dimensional irrotational ideal
fluid \cite{whitham74,johnson97,craik04,craik05} bounded below by a
flat wall and above by an evolving surface, $\eta(x,t)$.  Because the
flow is irrotational, there is a velocity potential $\phi$ such that
$\mathbf{u}=\nabla\phi$.  The restriction of $\phi$ to the free
surface is denoted $\varphi(x,t)=\phi(x,\eta(x,t),t)$.  The equations
of motion governing $\eta(x,t)$ and $\varphi(x,t)$ are
\begin{subequations}\label{eq:ww}
  \begin{align}
\label{eq:ww:1}
    \eta_t &= \phi_y - \eta_x\phi_x, \\[-3pt]
\label{eq:ww:2}
\varphi_t &= P\left[\phi_y\eta_t - \frac{1}{2}\phi_x^2 -
  \frac{1}{2}\phi_y^2 - g\eta +
  \frac{\sigma}{\rho}\partial_x\left(\frac{\eta_x}{\sqrt{1+\eta_x^2}}
  \right)\right].
  \end{align}
\end{subequations}
Here $g$ is the acceleration of gravity, $\rho$ is the fluid density,
$\sigma\ge0$ is the surface tension (possibly zero), and $P$ is the
$L^2$ projection to zero mean that annihilates constant functions,
\begin{equation}
  P = \opn{id} - P_0, \qquad
  P_0 f = \frac{1}{2\pi}\int_0^{2\pi} f(x)\,dx.
\end{equation}
This projection is not standard in (\ref{eq:ww:2}), but yields a
convenient convention for selecting the arbitrary additive constant in
the potential.  In fact, if the fluid has infinite depth and the mean
surface height is zero, $P$ has no effect in (\ref{eq:ww:2}) at the
PDE level, ignoring roundoff and discretization errors.
The velocity components $u=\phi_x$, $v=\phi_y$ on the right hand side
of (\ref{eq:ww}) are evaluated at the free surface to determine
$\eta_t$ and $\varphi_t$.  The system is closed by relating $\phi$ in
the fluid to $\eta$ and $\varphi$ on the surface as the solution of
Laplace's equation
\begin{subequations}\label{eq:dno}
\begin{align}
\label{eq:dno:1}
    \phi_{xx} + \phi_{yy} &= 0, & -\hh &< y < \eta, \\
\label{eq:dno:2}
    \phi_y &= 0, &  y &= -\hh, \\
\label{eq:dno:3}
    \phi &= \varphi, & y &= \eta,
\end{align}
\end{subequations}
where $\hh$ is the mean fluid depth (possibly infinite). We assume
$\eta(x,t)$ and $\mathbf{u}(x,y,t)$ are $2\pi$-periodic in
$x$. Applying a horizontal Galilean transformation if necessary, we
may also assume $\phi$ is $2\pi$-periodic in $x$.  We generally
assume $\hh=0$ in the finite depth case and absorb the mean fluid depth
into $\eta$ itself. This causes $-\eta(x)$ to be a reflection of the
free surface across the bottom boundary, which simplifies many
formulas in the boundary integral formulation below. The same strategy
can also be applied in the presence of a more general bottom topography
\cite{nachbin:04}.

Equation (\ref{eq:ww:1}) is a kinematic condition requiring that
particles on the surface remain there. Equation (\ref{eq:ww:2}) comes
from $\varphi_t = \phi_y\eta_t + \phi_t$ and the unsteady Bernoulli
equation, $\phi_t + \frac{1}{2}|\nabla\phi|^2 + gy + \frac{p}{\rho} =
c(t)$, where $c(t)$ is constant in space but otherwise arbitrary.
At the free surface, we assume the pressure jump across the interface
due to surface tension is proportional to curvature,
$p_0-p\vert_{y=\eta} = \sigma\kappa$.  The ambient pressure $p_0$ is
absorbed into the arbitrary function $c(t)$, which is chosen to
preserve the mean of $\varphi(x,t)$:
\begin{equation}\label{eq:c:def}
  c(t) = \frac{p_0}{\rho} + P_0\left[
    \eta_x\phi_x\phi_y + \frac{1}{2}\phi_x^2 - \frac{1}{2}\phi_y^2
    + g\eta - \frac{\sigma}{\rho}\partial_x\left(\frac{\eta_x}{
        \sqrt{1+\eta_x^2}}\right)
    \right].
\end{equation}
The advantage of this construction is that $\mathbf{u}=\nabla\phi$ is
time-periodic with period $T$ if and only if $\eta$ and $\varphi$ are
time-periodic with the same period.  Otherwise, $\varphi(x,T)$ could
differ from $\varphi(x,0)$ by a constant function without affecting
the periodicity of $\mathbf{u}$.

Details of our boundary integral formulation are given in
Appendix~\ref{sec:BI}.  Briefly, we identify $\mathbb{R}^2$ with
$\mathbb{C}$ and parametrize the free surface by
\begin{equation}\label{eq:zeta}
  \zeta(\alpha) = \xi(\alpha) + i\eta(\xi(\alpha)),
\end{equation}
where the change of variables $x=\xi(\alpha)$ allows for smooth mesh
refinement in regions of high curvature, and $t$ has been suppressed
in the notation.  We compute the Dirichlet-Neumann operator
\cite{craig:sulem:93},
\begin{equation}\label{eq:DNO:def}
  \mc G\varphi(x) = \sqrt{1+\eta'(x)^2}\,\, \der{\phi}{n}(x+i\eta(x)),
\end{equation}
which appears implicitly in the right hand side of (\ref{eq:ww})
through $\phi_x$ and $\phi_y$, in three steps.  First, we solve the
integral equation
\begin{equation}\label{eq:fred}
  \frac{1}{2}\mu(\alpha) + \frac{1}{2\pi}\int_0^{2\pi}
  [K_1(\alpha,\beta) + K_2(\alpha,\beta)]\mu(\beta)\,d\beta =
  \varphi(\xi(\alpha))
\end{equation}
for the dipole density, $\mu(\alpha)$, in terms of the (known)
Dirichlet data $\varphi(\xi(\alpha))$.  Formulas for $K_1$ and $K_2$
are given in (\ref{eq:K1K2}) below.  These kernels are smooth
functions (even at $\alpha=\beta$), so the integral is not singular;
see Appendix~\ref{sec:BI}.  Second, we differentiate $\mu(\alpha)$ to
obtain the vortex sheet strength, $\gamma(\alpha)=\mu'(\alpha)$.
Finally, we evaluate the normal derivative of $\phi$ at the free
surface via
\begin{equation}\label{eq:G}
  \mc G\varphi(\xi(\alpha)) = \frac{1}{|\xi'(\alpha)|}
  \left[\frac{1}{2}H\gamma(\alpha) + \frac{1}{2\pi}\int_0^{2\pi}
    [G_1(\alpha,\beta) + G_2(\alpha,\beta)]\gamma(\beta)\,d\beta
    \right].
\end{equation}
$G_1$ and $G_2$ are defined in (\ref{eq:K1K2}) below, and $H$ is the
Hilbert transform, which is diagonal in Fourier space with symbol
$\hat H_k=-i\opn{sgn}(k)$.  The only unbounded operation in this
procedure is the second step, in which $\gamma(\alpha)$ is obtained
from $\mu(\alpha)$ by taking a derivative.

Once $\mc G\varphi(x)$ is known, we compute $\phi_x$ and $\phi_y$ on
the boundary using
\begin{equation}\label{eq:uv:from:G}
  \begin{pmatrix} \phi_x \\ \phi_y \end{pmatrix} =
  \frac{1}{1+\eta'(x)^2}\begin{pmatrix}
    1 & -\eta'(x) \\ \eta'(x) & 1 \end{pmatrix}
  \begin{pmatrix}
    \varphi'(x) \\
    \mc G\varphi(x)
  \end{pmatrix},
\end{equation}
which allows us to evaluate (\ref{eq:ww:1}) and (\ref{eq:ww:2}) for
$\eta_t$ and $\varphi_t$.  Alternatively, one can write the
right hand side of (\ref{eq:ww}) directly in terms of $\varphi'(x)$
and $\mc G\varphi(x)$.

\subsection{GPU-accelerated time-stepping and quadruple precision}

\begin{figure}[b]
\begin{center}
\includegraphics[width=.8\linewidth]{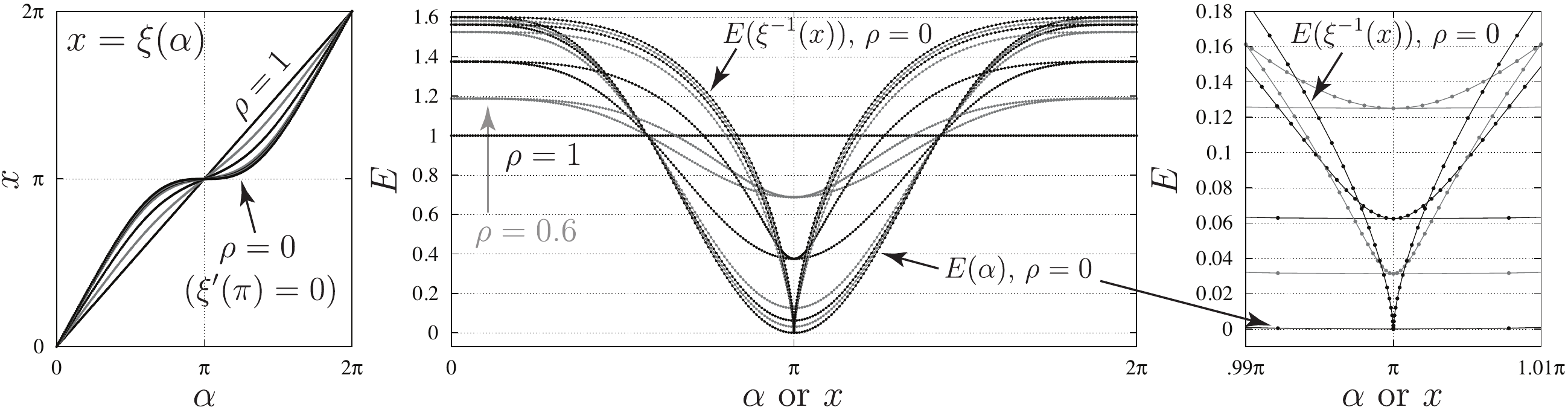}
\end{center}
\caption{\label{fig:spacing} Dependence of mesh spacing on the
  parameter $\rho$ (dropping the subscript $l$) in (\ref{eq:xi}), with
  mesh refinement near $x=\pi$.  (left) Plots of $x=\xi(\alpha)$ for
  $\rho=0.0$, $0.02$, $0.04$, $0.08$, $0.25$, $0.6$ and $1.0$.
  (center) $E(\alpha)=\partial\xi/\partial\alpha$ represents the grid
  spacing relative to uniform spacing.  Comparison of $E(\alpha)$ and
  $E(\xi^{-1}(x))$ shows how the grid points are re-distributed.
  (right) A magnified view near $\alpha=\pi$ shows that when $\rho$
  reaches 0, $\xi(\alpha)$ ceases to be a diffeomorphism and
  $E(\xi^{-1}(x))$ forms a cusp.  }
\end{figure}

Next we turn to the question of discretization.  Because we are
interested in studying large amplitude standing waves that develop
relatively sharp wave crests for brief periods of time, we discretize
space and time adaptively.  Time is divided into $\nu$ segments
$\theta_lT$, where $\theta_1 + \cdots + \theta_\nu = 1$ and $T$ is the
simulation time, usually an estimate of the period or quarter-period.
In the simulations reported here, $\nu$ ranges from 1 to 5 and each
$\theta_l$ was close to $1/\nu$ (within a factor of two).  On segment
$l$, we fix the number of (uniform) timesteps, $N_l$, the number of
spatial grid points, $M_l$, and the function
\begin{equation}\label{eq:xi}
  \xi_l(\alpha) = \int_0^\alpha E_l(\beta)\,d\beta, \quad
  E_l(\alpha) = \left\{\!\!\begin{array}{rl}
    1 - P\big[A_l\sin^4(\alpha/2)\big], & \text{to refine near } x=\pi \\[2pt]
    1 - P\big[A_l\cos^4(\alpha/2)\big], & \text{to refine near } x=0
  \end{array}\!\!\right\}, \quad
  A_l = \frac{8(1-\rho_l)}{5+3\rho_l},
\end{equation}
which controls the grid spacing in the change of variables
$x=\xi_l(\alpha)$; see Figure~\ref{fig:spacing}.
As before, $P$ projects out the mean. The parameter
$\rho_l$ lies in the range $0<\rho_l\le1$ and satisfies
\begin{equation}\label{eq:rho:def}
  \rho_l = \frac{\min\{E_l(0), E_l(\pi)\}}{\max\{E_l(0), E_l(\pi)\}}, \qquad
  \min\{E_l(0), E_l(\pi)\} = \frac{8\rho_l}{5+3\rho_l}, \qquad
  \max\{E_l(0), E_l(\pi)\} = \frac{8}{5+3\rho_l}.
\end{equation}
Note that $\rho_l=1$ corresponds to uniform spacing while $\rho_l=0$
corresponds to the singular limit where $\xi_l$ ceases to be a
diffeomorphism at one point. This approach takes advantage of the fact
that we can arrange in advance that the wave crests will form at $x=0$
and $x=\pi$, alternating between the two in time.  A more automated
approach would be to have the grid spacing evolve with the wave
profile, perhaps as a function of curvature, rather than asking the
user to specify the change of variables.  We did not experiment with
this idea since our approach also allows the number of grid points to
increase in time, which would be complicated in an automated approach.
We always set $\rho_1=1$ so that $x=\alpha$ on the first segment.
Respacing the grid from segment $l$ to $l+1$ boils down to
interpolating $\eta$ and $\varphi$ to obtain values on the new mesh,
e.g.~$\eta\circ\xi_{l+1}(\alpha_j) = \eta\circ\xi_l(
\xi_l^{-1}\circ\xi_{l+1}(\alpha_j))$, $\alpha_j=2\pi j/M_l$, which is
straightforward by Newton's method.  To be safe, we avoid refining the
mesh in one region at the expense of another; thus, if $\rho_{l+1}<
\rho_l$, we also require $(M_{l+1}/M_l)\ge(5+3\rho_l)/(5+3\rho_{l+1})$
so that the grid spacing decreases throughout the interval, but more
so in the region where the wave crest is forming.

Since the evolution equations are not stiff unless the surface tension
is large, high order explicit time-stepping schemes work well.  For
each Runge-Kutta stage within a timestep on a given segment~$l$, the
integral equation (\ref{eq:fred}) is solved by collocation using
uniformly spaced grid points $\alpha_j=2\pi j/M_l$ and the (spectrally
accurate) trapezoidal rule,
\begin{equation}
  \frac{1}{2\pi}\int_0^{2\pi}K(\alpha_i,\beta)\mu(\beta)\,d\beta \approx
  \frac{1}{M_l} \sum_{j=0}^{M_l-1} K(\alpha_i,\alpha_j)\mu(\alpha_j).
\end{equation}
The matrices
$K_{ij}=K(\alpha_i,\alpha_j)/M_l$ and
$G_{ij}=G(\alpha_i,\alpha_j)/M_l$ that represent the discretized
integral operators in (\ref{eq:fred}) and (\ref{eq:G}) are computed
simultaneously and in parallel. The formulas are
$K(\alpha,\beta)=K_1(\alpha,\beta)+K_2(\alpha,\beta)$ and
$G(\alpha,\beta)=G_1(\alpha,\beta)+G_2(\alpha,\beta)$ with
\begin{equation}\label{eq:K1K2}
\begin{aligned}
  &K_1 = \im\left\{\frac{\zeta'(\beta)}{2}
  \cot\frac{\zeta(\alpha) - \zeta(\beta)}{2}
  - \frac{1}{2}\cot\frac{\alpha-\beta}{2}\right\}, \quad
  K_2 = \im\left\{\frac{\bar\zeta'(\beta)}{2}\cot
  \frac{\zeta(\alpha)-\bar\zeta(\beta)}{2}\right\}, \\
  & G_1 = \re\left\{\frac{\zeta'(\alpha)}{2}
  \cot\frac{\zeta(\alpha) - \zeta(\beta)}{2}
  - \frac{1}{2}\cot\frac{\alpha-\beta}{2}\right\}, \quad
  G_2 = \re\left\{\frac{\zeta'(\alpha)}{2}\cot
  \frac{\zeta(\alpha)-\bar\zeta(\beta)}{2}\right\}.
\end{aligned}
\end{equation}
As explained in Appendix~\ref{sec:BI}, these kernels have been
regularized.  Indeed, $K_1(\alpha,\beta)$ and $G_1(\alpha,\beta)$ are
continuous at $\beta=\alpha$ if we define $K_1 =
-\im\{\zeta''(\alpha)/[2\zeta'(\alpha)]\}$ and
$G_1=\re\{\zeta''(\alpha)/[2\zeta'(\alpha)]\}$.  These formulas are
used when computing the diagonal entries $K_{ii}$ and $G_{ii}$.  The
terms $\cot((\alpha_i-\alpha_j)/2)$ in (\ref{eq:K1K2}) are computed
once and for all at the start. If the fluid depth is infinite, $K_2$
and $G_2$ are omitted.  GMRES is used to solve (\ref{eq:fred}) for
$\mu$, which consistently takes 4-30 iterations to reach machine
precision (independent of problem size).  In quadruple precision, the
typical range is 9-36 GMRES iterations.  The FFT is used to compute
$\mu'$ and $H\gamma$ in (\ref{eq:G}), as well as $\zeta'$, $\zeta''$,
$\eta'$, and $\varphi'$.

We wrote 3 versions of the code, which differ only in how the matrices
$K$ and $G$ are computed.  The simplest version uses openMP
\emph{parallel for} loops to distribute the work among all available
threads.  The most complicated version is parallelized using MPI and
scalapack. In this case, the matrices $K$ and $G$ are stored in
block-cyclic layout \cite{Lim96} across the processors, and each
processor computes only the matrix entries it is responsible for. The
fastest version of the code is parallelized on a GPU in the cuda
programming language.  First, the CPU sends the GPU the vector
$\zeta(\alpha_j)$, which holds $M_l$ complex numbers.  Next, the GPU
computes the matrices $K$ and $G$ and stores them in device memory.
Finally, in the GMRES iteration, Krylov vectors are sent to the GPU,
which applies the matrix $K$ and returns the result as a vector.
After the last Krylov iteration, the device also applies $G$ to $\mu$
to help compute $\mc{G}\varphi$ in (\ref{eq:G}).  Thus, communication
with the GPU involves passing vectors of length $M_l$, while
$O(M_l^2)$ flops must be performed on each vector passed in.  As a
result, communication does not pose a computational bottleneck, and
the device operates at near 100\% efficiency.  We remark that the
formula
\begin{equation*}
  \cot\frac{x+iy}{2} = \begin{cases}
    [\cos(x)+\cosh(y)]\big/[\sin(x)+i\sinh(y)], & \cos(x)\ge 0, \\
    [\sin(x)-i\sinh(y)]\big/[\cosh(y)-\cos(x)], & \cos(x)<0
  \end{cases}
\end{equation*}
is relatively expensive to evaluate. Thus, it pays to compute $K$ and
$G$ simultaneously (to re-use $\sin$, $\cos$, $\sinh$, $\cosh$
results), and to actually store the matrices in device memory rather
than re-compute the matrix entries each time a matrix-vector product
is required.

In double-precision, we evolve (\ref{eq:ww}) using Dormand and
Prince's DOP853 scheme \cite{hairer:I}.  This is a 13 stage, 8th
order, ``first same as last'' Runge-Kutta method, so the effective
cost of each step is 12 function evaluations.  We apply the 36th order
filter described in \cite{hou:li:07} to the right hand side of (1e)
and (1f) each time they are evaluated in the Runge-Kutta procedure,
and to the solution itself at the end of each time-step.  This filter
consists of multiplying the $k$th Fourier mode by
\begin{equation}\label{eq:filter}
  \exp\left[-36\big(|k|/k_\text{max}\big)^{36}\right], \qquad
  k_\text{max} = M/2,
%  e^{-36\rho_k^{36}}, \qquad \rho_k = |k|/(M/2),
\end{equation}
which allows the highest-frequency Fourier modes to remain non-zero
(to help resolve the solution) while still suppressing aliasing
errors.  To achieve truncation errors of order $10^{-30}$ in
quadruple-precision, the 8th order method requires too many
timesteps. Through trial and error, we found that a 15th order
spectral deferred correction (SDC) method \cite{dutt,huang,minion} is
the most efficient scheme for achieving this level of accuracy.  Our
GPU implementation of quadruple precision arithmetic will be discussed
briefly in Section~\ref{sec:trust}. The variant of SDC that we use in
this paper employs eight Radau IIa quadrature nodes
\cite{hairer:I}. The initial values at the nodes are obtained via
fourth order Runge-Kutta.  Ten correction sweeps are then performed to
improve the solution to $O(h^{15})$ accuracy at the quadrature nodes.
We use pure Picard corrections instead of the more standard
forward-Euler corrections as they have slightly better stability
properties.  The final integration step yields a local truncation
error of $O(h^{16})$; hence, the method is 15th order.  See
\cite{sdc:picard} for more information about this variant of the SDC
method and its properties.  If one wished to go beyond
quadruple-precision arithmetic, it is straightforward to increase the
order of the time-stepping scheme accordingly.  We did not investigate
the use of symplectic integrators since our approach already conserves
energy to 12-16 digits of accuracy in double precision, and 24-32
digits in quadruple precision.

\subsection{Translational and time-reversal symmetry}
\label{sec:sym}

In this paper, we restrict attention to symmetric standing waves of
the type studied in \cite{rayleigh1876, penney:52, tadjbakhsh,
  concus:62, mercer:92, mercer:94, schultz, ioualalen:03}. For these
waves, it is only necessary to evolve the solution over a quarter
period.  Indeed, if at some time $T/4$ the fluid comes to rest
($\varphi\equiv0$), a time-reversal argument shows that the solution
will evolve back to the initial state at $T/2$ with the sign of
$\varphi$ reversed.  More precisely, the condition $\varphi(x,T/4)=0$
implies that $\eta(x,T/2)=\eta(x,0)$ and $\varphi(x,T/2)=
-\varphi(x,0)$.  Now suppose that, upon translation by $\pi$,
$\eta(x,0)$ remains invariant while $\varphi(x,0)$ changes sign.  Then
we see that $\eta_1(x,t)=\eta(x+\pi,T/2+t)$ and
$\varphi_1(x,t)=\varphi(x+\pi,T/2+t)$ are solutions of (\ref{eq:ww})
with initial conditions
\begin{align*}
\eta_1(x,0)&=\eta(x+\pi,T/2)=\eta(x+\pi,0)=\eta(x,0),\\
\varphi_1(x,0)&=\varphi(x+\pi,T/2)=-\varphi(x+\pi,0)=\varphi(x,0).
\end{align*}
Therefore, $\eta_1=\eta$, $\varphi_1=\varphi$, and
\begin{align*}
  \eta(x,T) &= \eta_1(x-\pi,T/2) = \eta(x-\pi,T/2) = \eta(x-\pi,0) = \eta(x,0), \\
  \varphi(x,T) &= \varphi_1(x-\pi,T/2) = \varphi(x-\pi,T/2) =
  -\varphi(x-\pi,0) = \varphi(x,0).
\end{align*}
Hence, $\eta$ and $\varphi$ are time-periodic with period $T$.  It is
natural to expect standing waves to have even symmetry when the origin
is placed at a crest or trough and the fluid comes to rest.
% (i.e.~at $t=T/4$ and $t=3T/4$).
This assumption implies that $\eta$ and $\varphi$ will remain even
functions for all time since $\eta_t$ and $\varphi_t$ in (\ref{eq:ww})
are even whenever $\eta$ and $\varphi$ are. Under all these
assumptions, the evolution of $\eta$ and $\varphi$ from $T/2$ to $T$
is a mirror image (about $x=\frac{\pi}{2}$ or $x=\frac{3\pi}{2}$) of
the evolution from $0$ to $T/2$.

Once the initial conditions and
period are found using symmetry to accelerate the search for
time-periodic solutions, we double-check that the numerical solution
evolved from $0$ to $T$ is indeed time-periodic.  Mercer and Roberts
exploited similar symmetries in their numerical computations
\cite{mercer:92,mercer:94}.

\section{Overdetermined shooting methods}
\label{sec:od:shoot}

As discussed in the introduction, two-point boundary value problems
governed by partial differential equations must be discretized before
solving them numerically.  However, truncation errors lead to loss of
accuracy in the highest-frequency modes of the numerical solution,
which can cause difficulty for the convergence of shooting methods.
We will see below that robustness can be achieved by
posing these problems as overdetermined nonlinear systems.

In Section~\ref{sec:n:ls}, we define two objective functions with the
property that driving them to zero is equivalent to finding a
time-periodic standing wave. One of the objective functions exploits
the symmetry discussed above to reduce the simulation time by a factor
of 4. The other is more robust as it naturally penalizes
high-frequency Fourier modes of the initial conditions.  Both
objective functions use symmetry to reduce the number of unknowns and
eliminate phase shifts of the standing waves in space and time. The
problem is overdetermined because the highest-frequency Fourier modes
are constrained to be zero initially but not at the final time.  Also,
because $T/4$ often corresponds to a sharply crested wave profile,
there are more active Fourier modes in the solution at that time than
at $t=0$. By refining the mesh adaptively, we include all of these
active modes in the objective functions, making them more
overdetermined. The idea that the underlying dynamics of standing
water waves is lower-dimensional than predicted by counting active
Fourier modes has recently been explored by Williams,
\emph{et al.}~\cite{pod}.

In Sections~\ref{sec:acm} and~\ref{sec:trust}, we describe two methods
for solving the resulting nonlinear least squares problem. The first
is the Adjoint Continuation Method \cite{benj1,benj2,vtxs1,lasers}, in
which the gradient of the objective function is computed by solving an
adjoint PDE and the BFGS algorithm \cite{bfgs, nocedal} is used for
the minimization. The second is a trust-region approach in which the
Jacobian is computed by solving the variational equation in parallel
with multiple right-hand sides. This allows the work of computing the
Dirichlet-Neumann operator to be shared across all the columns of the
Jacobian. We also discuss implementation issues in quadruple precision
on a GPU.

\subsection{Nonlinear least squares formulation}
\label{sec:n:ls}

In the symmetric standing wave case considered here, we assume the
initial conditions are even functions satisfying
$\eta(x+\pi,0)=\eta(x,0)$ and $\varphi(x+\pi,0)=-\varphi(x,0)$.  In
Fourier space, they take the form
\begin{equation}\label{eq:init:stand}
  \begin{aligned}
    \hat\eta_k(0) &= c_{|k|}, \qquad (k=\pm2,\pm4,\pm6,\dots\;;\; |k|\le n), \\
    \hat\varphi_k(0) &= c_{|k|}, \qquad (k=\pm1,\pm3,\pm5,\dots\;;\; |k|\le n),
  \end{aligned}
\end{equation}
where $c_1,\dots,c_n$ are real numbers, and all other Fourier modes of
the initial conditions are set to zero. (In the finite depth
case, we also set $\hat\eta_0=\hh$, the mean fluid depth.) Here $n$
is taken to be somewhat smaller than $M_1$, e.g.~$n\approx
\frac{1}{3}M_1$, where $M_1$ is the number of spatial grid points used
during the first $N_1$ timesteps. (Recall that subscripts on $M$ and
$N$ refer to mesh refinement sub-intervals.) Note that high-frequency
Fourier modes of the initial condition are zero-padded to improve
resolution of the first $n$ Fourier modes.

In addition to the Fourier modes of the initial condition, the period
of the solution is unknown. We add a zeroth component to $c$ to
represent the period:
\begin{equation}\label{eq:T:c0}
  T = c_0.
\end{equation}
Our goal is to find $c\in\mathbb{R}^{n+1}$ such that $\varphi(x,T/4)=0$.
We therefore define the objective function
\begin{equation}\label{eq:f:phi}
  f(c) = \frac{1}{2}r(c)^Tr(c) \approx
  \frac{1}{4\pi}\int_0^{2\pi}
  \varphi(x,T/4)^2\,dx,
  \qquad
  r_i = \varphi(\xi_\nu(\alpha_i),T/4)\sqrt{E_\nu(\alpha_i)/M_\nu},
\end{equation}
where $\nu$ is the index of the final sub-interval in the mesh
refinement strategy and the square root is a quadrature weight to
approximate the integral via the trapezoidal rule after the change of
variables $x=\xi_\nu(\alpha)$, $dx=E_\nu(\alpha)\,d\alpha$. Note that
$r\in\mathbb{R}^m$ with $m=M_\nu$, which is usually several times
larger than $n$, the number of non-zero initial conditions. The
numerical solution is not sensitive to the choice of $m$ and $n$ as
long as enough zero-padding is included in the initial condition to
resolve the highest frequency Fourier modes. This will be confirmed
in Section~\ref{sec:breakdown} through mesh-refinement studies and
comparison with quadruple-precision computations.

One can also use an objective function that measures deviation from
time-periodicity directly:
\begin{equation}\label{eq:f:naive}
  f(c) \approx
    \frac{1}{4\pi}\int_0^{2\pi}
    \big[ \eta(x,T) - \eta(x,0) \big]^2 +
    \big[ \varphi(x,T) - \varphi(x,0) \big]^2\,dx.
\end{equation}
When the underlying PDE is stiff (e.g.~for the Benjamin-Ono
\cite{benj1,benj2} or KdV equations), an objective function of the
form (\ref{eq:f:naive}) has a key advantage over (\ref{eq:f:phi}).
For stiff problems, semi-implicit time-stepping methods are used in
order to take reasonably large time-steps. Such methods damp high-frequency
modes of the initial condition. This causes these modes to
have little effect on an objective function of the form
(\ref{eq:f:phi}); thus, the Jacobian $J_{ij}=\partial r_i/\partial
c_j$ can be poorly conditioned if the shooting method attempts to
solve for too many modes.  By contrast, when implemented via
(\ref{eq:f:naive}), the initial conditions of high-frequency modes are
heavily penalized for deviating from the damped values at time $T$.
As a result, the Jacobian does not suffer from rank deficiency, and
high-frequency modes do not drift far from zero unless doing so is
helpful.  Since the water wave is not stiff, we use explicit schemes
that do not significantly damp high-frequency modes; therefore, the
computational advantage of evolving over a quarter-period
outweigh any robustness advantage of using (\ref{eq:f:naive}).

We used symmetry to reduce the number of unknown initial conditions in
(\ref{eq:init:stand}).  This has the added benefit of selecting the
spatial and temporal phase of each solution in a systematic manner.
In problems where the symmetries of the solution are not known in
advance, or to search for symmetry-breaking bifurcations, one can
revert to the approach described in \cite{benj1}, where both real and
imaginary parts of the leading Fourier modes of the initial condition
were computed in the search for time-periodic solutions.  To eliminate
spatial and temporal phase shifts, one of the Fourier modes was
constrained to be real and its time derivative was required to be
imaginary.  Constraining the time-derivative of a mode is most easily
done with a penalty function \cite{benj1}.  Alternatively, if two
modes are constrained to be real and their time-derivatives are left
arbitrary, it is easier to remove their imaginary parts from the
search space than to use a penalty function.

Once phase shifts have been eliminated, the families of time-periodic
solutions we have found appear to sweep out two-parameter families of
solutions.  To compute a solution in a family, we specify the mean
depth and the value of one of the $c_k$ in (\ref{eq:init:stand}) or
(\ref{eq:T:c0}) and solve for the other $c_j$ to minimize the
objective function.  If $f$ is reduced below a specified threshold
(typically $10^{-26}$ in double-precision or $10^{-52}$ in quadruple
precision), we consider the solution to be time-periodic. If $f$
reaches a local minimum that is higher than the specified threshold,
we either (1) refine the mesh, increase $n$, and try again; (2) choose
a different value of $c_k$ closer to the previous successful value; or
(3) change the index $k$ specifying which Fourier mode is used as a
bifurcation parameter. Switching to a different $k$ is often useful
when tracking a fold in the bifurcation curve.  Since
$c\in\mathbb{R}^{n+1}$ and one parameter has been frozen, $f$ is
effectively a function of $n$ variables.

We note that once $n$ and the mesh parameters $\nu$, $\theta_l$,
$A_l$, $M_l$ and $N_l$ are chosen, $f(c)$ is a smooth function that
can be minimized using a variety of optimization techniques. Small
divisors come into play when deciding whether $f$ would really
converge to zero in the mesh refinement limit (with ever increasing
numerical precision).  The answer may depend on whether the bifurcation
parameters ($\hat\eta_0$ and either $\eta(a,0)$ or one of the $c_k$)
are allowed to vary within the tolerance of the current roundoff
threshold each time the mesh is refined and the floating point
precision is increased. While it is likely that small divisors prevent
the existence of smooth families of exact solutions, exceedingly
accurate approximate solutions do appear to sweep out smooth families,
with occasional disconnections in the bifurcation curves due to
resonance.

\subsection{Adjoint continuation method}
\label{sec:acm}

Having recast the shooting method as an overdetermined nonlinear least
squares problem, we must now minimize the functional $f$ in
(\ref{eq:f:phi}) or (\ref{eq:f:naive}). The first approach we tried
was the adjoint continuation method (ACM)
developed by Ambrose and Wilkening to study time-periodic solutions of
the Benjamin-Ono equation \cite{benj1,benj2} and the vortex sheet with
surface tension \cite{vtxs1}. The method has also been used by
Williams \emph{et al.}~to study the stability transition from single-pulse
to multi-pulse dynamics in a mode-locked laser system \cite{lasers}.

The idea of the ACM is to compute the gradient of $f$ with respect to
the initial conditions by solving an adjoint PDE, and then minimize
$f$ using the BFGS method \cite{bfgs,nocedal}.  BFGS is a
quasi-Newton algorithm that builds an approximate (inverse) Hessian
matrix from the sequence of gradient vectors it encounters on
successive line-searches.
In more detail, let $q=(\eta,\varphi)$ and denote the system
(\ref{eq:ww}) abstractly by
\begin{equation}\label{eq:qt}
  q_t = F(q), \qquad q(x,0)=q_0(x).
\end{equation}
We define the inner product
\begin{equation}\label{eq:inner:prod}
  \langle q_1,q_2\rangle=\frac{1}{2\pi}\int_0^{2\pi}
  \left[\eta_1(x)\eta_2(x)+\varphi_1(x)\varphi_2(x)\right]dx
\end{equation}
so that $f$ in (\ref{eq:f:phi}), written now as a function of the
initial conditions and proposed period, which themselves depend on $c$
via (\ref{eq:init:stand}) and (\ref{eq:T:c0}), takes the form
\begin{equation}\label{eq:f:phi:again}
  f(q_0,T) = \frac{1}{2}\|\,\big(0,\varphi(\cdot,T/4)\big)\,\|^2 =
  \frac{1}{4\pi} \int_0^{2\pi} \varphi(x,T/4)^2\,dx,
\end{equation}
where $q=(\eta,\varphi)$ solves (\ref{eq:qt}).  The case with $f$ of
the form (\ref{eq:f:naive}) is similar, so we omit details here.
In the course of minimizing $f$, the BFGS algorithm will repeatedly
query the user to evaluate both $f(c)$ and its gradient $\nabla_c
f(c)$ at a sequence of points $c\in\mathbb{R}^{n+1}$.  The $T$
derivative, $\partial f/\partial c_0$, is easily obtained by
evaluating
\begin{equation}\label{eq:dfdT}
  \der{f}{T} = \frac{1}{8\pi}\int_0^{2\pi}
  \varphi(x,T/4)\,\varphi_t(x,T/4)\,dx
\end{equation}
using the trapezoidal rule after changing variables,
$x=\xi_\nu(\alpha)$, $dx=E_\nu(\alpha)\,d\alpha$. Note that
$\varphi(\cdot,T/4)$ and $\varphi_t(\cdot,T/4)$ are already known by
solving (\ref{eq:ww}).
One way to compute the other components of $\nabla_c f$, say $\partial
f/\partial c_k$, would be to solve the variational equation, (written
abstractly here and explicitly in Appendix~\ref{sec:adjoint})
\begin{equation}\label{eq:dot}
  \dot{q}_t=D F(q(\cdot,t))\dot{q}, \qquad
  \dot q(x,0) = \dot q_0(x)
\end{equation}
with initial conditions
\begin{equation}\label{eq:dot:ic}
  \dot q_0(x) = \begin{cases}
    (e^{ikx}+e^{-ikx},0), & k=1,3,5,\dots \\
    (0,e^{ikx}+e^{-ikx}), & k=2,4,6,\dots
  \end{cases}
\end{equation}
to obtain
\begin{equation}\label{eq:f:dot1}
  \der{f}{c_k} = 
  \dot{f} = \left.\frac{d}{d\veps}\right|_{\veps=0}
  f(q_0+\veps \dot{q}_0,T)=
  \big\langle\, \big(0,\varphi(\cdot,T/4)\big)\,,
  \big(0,\dot{\varphi}(\cdot,T/4)\big)\,\big\rangle.
\end{equation}
Note that a dot denotes a directional derivative with respect to the
initial condition, not a time-derivative.  To avoid the expense of
solving (\ref{eq:dot}) repeatedly for each value of $k$, we solve a
single adjoint PDE to find $\delta f/\delta q_0$ such that
$\dot f = \langle\, \delta f/\delta q_0\,,\,\dot q_0 \,\rangle$.
From (\ref{eq:f:dot1}), we have
\begin{equation}\label{eq:f:dot2}
  \dot{f} = \big\langle\,\big(0,\varphi(\cdot,T/4)\big)\,,
  \big( \dot\eta(\cdot,T/4),\dot{\varphi}(\cdot,T/4)\big)\,
  \big\rangle = \langle\, \tilde q_0\,,\dot q(\cdot,T/4)\,\rangle,
\end{equation}
where we have defined $\tilde q_0 = (\tilde\eta_0,\tilde\varphi_0)$
with $\tilde\eta_0 = 0$ and $\tilde\varphi_0 = \varphi(\cdot,T/4)$.
Note that replacing 0 by $\dot\eta(\cdot,T/4)$ did not affect the
inner product.  Next we observe that the solution $\tilde q(x,s)$ of
the adjoint equation
\begin{equation}\label{eq:adj}
  \tilde{q}_s=D F(q(\cdot,T/4-s))^*\tilde{q}, \qquad
  \tilde q(\cdot,0) = \tilde q_0,
\end{equation}
which evolves backward in time ($s=T/4-t$), has the property that
\begin{equation}
  \langle\tilde{q}(\cdot,T/4-t),\dot{q}(\cdot,t)\rangle=\text{const}.
\end{equation}
Setting $t=T/4$ shows that this constant is actually $\dot
f$.  Setting $t=0$ gives the form we want:
\begin{equation}
  \dot f = \langle\, \delta f/\delta q_0\,,\,\dot q_0 \,\rangle, \qquad
  \derr{f}{q_0} = \tilde q(\cdot,T/4).
\end{equation}
From (\ref{eq:dot:ic}), we obtain
\begin{equation}\label{eq:nabla:f}
  \der{f}{c_k} = \left\{\begin{array}{cc}
    2\re\big\{\wdg{\tilde\eta}_k(T/4)\big\}, & k = 1,3,5,\dots \\
    2\re\big\{\wdg{\tilde\varphi}_k(T/4)\big\}, & k = 2,4,6,\dots
  \end{array}\right\}.
\end{equation}
Together with (\ref{eq:dfdT}), this gives all the components of
$\nabla_c f$ at once.  Explicit formulas for the linearized and
adjoint equations (\ref{eq:dot}) and (\ref{eq:adj}) are derived in
Appendix~\ref{sec:adjoint}.

Like (\ref{eq:dot}), the adjoint equation (\ref{eq:adj}) is linear,
but non-autonomous, due to the presence of the solution $q(t)$ of
(\ref{eq:qt}) in the equation.  In the BFGS method, the gradient is
always called immediately after computing the function value; thus, if
$q(t)$ and $q_t(t)$ are stored in memory at each timestep in the
forward solve, they are available in the adjoint solve at intermediate
Runge-Kutta steps through cubic Hermite interpolation. We actually use
dense output formulas \cite{shampine86,hairer:I} for the 5th and 8th order
Dormand-Prince schemes since cubic Hermite interpolation limits the
accuracy of the adjoint solve to 4th order, but the idea is the same.
If there is insufficient memory to store the solution at every
timestep, we store the solution at equally spaced mile-markers and
re-compute $q$ between them when $\tilde q$ reaches that region.
Thus, $\nabla f$ can be computed in approximately
the same amount of time as $f$ itself, or twice the time if
mile-markers are used.

It is worth mentioning that, when discretized, the values of
$\eta$ and $\varphi$ are stored on a non-uniformly spaced grid
for each segment $l\in\{2,\dots,\nu\}$ in the mesh-refinement strategy.
The adjoint variables $\tilde\eta$, $\tilde\varphi$ are stored at the
same mesh points, and are initialized by
\begin{equation*}
  \tilde\eta_0\circ\xi_\nu(\alpha_i)=0, \qquad
  \tilde\varphi_0\circ\xi_\nu(\alpha_i) = \varphi(\xi_\nu(\alpha_i),T/4),
\end{equation*}
with no additional weight factors needed.
This works because the inner product (\ref{eq:inner:prod}) is defined
with respect to $x$ rather than $\alpha$, and the change of variables
has been accounted for by the factor $\sqrt{E_\nu(\alpha_i)/M_\nu}$ in
the formula (\ref{eq:f:phi}) for~$f$.

\subsection{Trust-region shooting method}
\label{sec:trust}

While the ACM method gives an efficient way of computing the gradient
of $f$, it takes many line-search iterations to build up an accurate
approximation of the Hessian of $f$.
This misses a key opportunity for parallelism and re-use of data that
can be exploited if we switch from the BFGS framework to a
Levenberg-Marquardt approach \cite{nocedal}. Instead of solving the
adjoint equation (\ref{eq:adj}) to compute $\nabla f$ efficiently, we
solve the variational equation (\ref{eq:dot}) with multiple right-hand
sides to compute all the columns of the Jacobian simultaneously.  From
(\ref{eq:f:phi}), we see that
\begin{equation}\label{eq:Jik}
  J_{ik} = \der{r_i}{c_k} = \begin{cases}
    \varphi_t(\xi_\nu(\alpha_i),T/4)\sqrt{E_\nu(\alpha_i)/M_\nu}, & k=0, \\
    \dot\varphi(\xi_\nu(\alpha_i),T/4)\sqrt{E_\nu(\alpha_i)/M_\nu}, & k\ge1,
  \end{cases}
\end{equation}
where $\dot q_0$ is initialized as in (\ref{eq:dot:ic}) for $k\ge1$.  We
avoid the need to store $q$ at every timestep (or at mile-markers) by
evolving $q$ along with $\dot q$ rather than interpolating $q$:
\begin{equation}\label{eq:q:qdot}
  \der{}{t}
  \begin{pmatrix} q \\ \dot q \end{pmatrix} =
  \begin{pmatrix} F(q) \\ DF(q)\dot q \end{pmatrix}, \qquad
  \begin{aligned}
    q(0) &= q_0 = (\eta_0,\varphi_0), \\
    \dot q(0) &= \dot q_0 = \partial q_0/\partial c_k.
  \end{aligned}
\end{equation}
In practice, we replace $\dot q$ in (\ref{eq:q:qdot}) by the matrix
$\dot Q=[\dot q_{(k=1)},\dots,\dot q_{(k=n)}]$ to compute all the
columns of $J$ (besides $k=0$) at once.
The linearized equations (\ref{eq:lin2}) involve the same
Dirichlet-to-Neumann operator as the nonlinear equations
(\ref{eq:ww}), so the matrices $K$ and $G$ in (\ref{eq:fred}) and
(\ref{eq:G}) only have to be computed once to evolve the entire matrix
$\dot Q$ through a Runge-Kutta stage.  Moreover, the linear algebra
involved can be implemented at level 3 BLAS speed.  For large
problems, we perform an LU-factorization of $K$, the cost of which is
made up for many times over by replacing GMRES iterations with a
single back-solve stage for each right-hand side.  In the GPU version
of the code, all the linear algebra involving $K$ and $G$ is performed
on the device (using the CULA library). As before, communication with
the device is minimal in comparison to the computational work
performed there.

We emphasize that the main advantage of solving linearized equations
is that the same DNO operator is used for each column of $\dot Q$ in a
given Runge-Kutta stage.  This opportunity is lost in the simpler
approach of approximating $J$ through finite differences by evolving
(\ref{eq:qt}) repeatedly, with initial conditions perturbed in each
coordinate direction:
\begin{equation}\label{eq:J:fd}
  J_{ik}\approx\frac{r_i(c+\veps e_k)-r_i(c)}{\veps},
  \qquad e_k = (0,\dots,0,1,0,\dots,0)^T\in\mathbb{R}^{n+1}.
\end{equation}
Thus, while finite differences can also be parallelized efficiently by
evolving
these solutions independently, the matrices $K$ and $G$ will be
computed $n$ times more often in the finite difference approach, and
most of the linear algebra will drop from running at level 3 BLAS
speed to level~2.  Details of our Levenberg-Marquardt implementation
are given in Appendix~\ref{sec:levmar}, where we discuss how to re-use
the Jacobian several times rather than re-computing it each time a
step is accepted.

The CULA and LAPACK libraries could not be used for
quadruple precision calculations, and we did not try FFTW in that
mode.  Instead, we used custom FFT and linear algebra libraries
(written by Wilkening) for this purpose.
However, for the GPU, we did not have any previous code to build on.
Our solution was to write a block version of matrix-matrix
multiplication in CUDA to compute residuals in quadruple precision,
then use iterative refinement to solve for the corrections in
double-precision, using the CULA library.  Although quadruple
precision is not native on any current GPU, we found M. Lu's
\emph{gqd} package \cite{gqd}, which is a CUDA implementation of
Bailey's \emph{qd} package \cite{qd}, to be quite fast.  Our code
is written so that the floating point type can be changed through a
simple \emph{typedef} in a header file. This is possible in C++ by
overloading function names and operators to call the appropriate
versions of routines based on the argument types.

\section{Numerical results}
\label{sec:results}

This section is organized as follows: In Section~\ref{sec:unit}, we
use the Adjoint Continuation Method to study standing waves of
wavelength $2\pi$ in water of uniform depth $h=1$.  Several
disconnections in the bifurcation curves are encountered, which are
shown to correspond physically to higher-frequency standing waves
superposed (nonlinearly) on the low-frequency carrier wave.  In
Section~\ref{sec:nuc}, we use the trust-region approach to study a
nucleation event in which isolated large-amplitude solutions, and
closed loops of such solutions, suddenly exist for depths below a
threshold value.  This gives a new mechanism for the creation of
additional branches of solutions (besides harmonic resonance
\cite{mercer:94, smith:roberts:99}).  In Section~\ref{sec:bif:degen},
we study a ``Wilton ripple'' phenomenon \cite{vandenBroeck:84,
  chen:saffman:79, bryant:stiassnie:94, bridges:standing:3d,
  akers:wilton} in which a pair of ``mixed mode'' solutions bifurcate
along side the ``pure mode'' solutions at a critical depth. Our
numerical solutions are accurate enough to identify the leading terms
in the asymptotic expansion of these mixed mode solutions. Following
the mixed-mode branches via numerical continuation reveals that they
meet up with the pure mode branches again at large amplitude. We also
study how this degenerate bifurcation splits when the fluid depth is
perturbed. In Section~\ref{sec:breakdown}, we study what goes wrong in
the Penney and Price conjecture, which predicts that the limiting
standing wave of extreme form will develop sharp 90 degree corner
angles at the wave crests.  We also discuss energy conservation, decay
of Fourier modes, and validation of accuracy.  In
Section~\ref{sec:shallow}, we study collisions of counter-propagating
solitary water waves that are elastic in the sense that the background
radiation is identical before and after the collision.  In
Section~\ref{sec:surf}, we study time-periodic gravity-capillary waves
of the type studied by Concus \cite{concus:62} and Vanden-Broeck
\cite{vandenBroeck:84} using perturbation theory. Finally, in
Section~\ref{sec:perform}, we compare the performance of the
algorithms on a variety of parallel machines, using MINPACK as a
benchmark for solving nonlinear least squares problems.

\subsection{Standing waves of unit depth}
\label{sec:unit}

We begin by computing a family of symmetric standing waves with mean
fluid depth $\hat\eta_0 = \hh = 1$ and zero surface tension.  The
linearized equations about a flat rest state are
\begin{equation}\label{eq:lin}
 \dot\eta_t = \mc{G}\dot\varphi, \qquad
 \dot\varphi_t = P[-g\dot\eta], \qquad
 \Big(\mc{G}\big[e^{ikx}\big] = \big[k\tanh k\hh\big]e^{ikx}\Big).
\end{equation}
Thus, the linearized problem has standing wave solutions of the
form
\begin{equation}\label{eq:lin:soln}
  \dot\eta = A\sin\omega t \cos kx, \quad
  \dot\varphi = B\cos\omega t \cos kx, \quad
  \omega^2 = kg\tanh k\hh, \quad
  A/B = \sqrt{(k/g)\tanh kh}.
\end{equation}
Setting $h=1$, $g=1$, $k=1$, these solutions have period
$T=2\pi/\omega \approx 7.19976$.  Here $B$ is a
free parameter controlling the amplitude, and $A$ is determined by
$A/B=\sqrt{\tanh 1}$.

\begin{figure}[b]
\includegraphics[width=\linewidth]{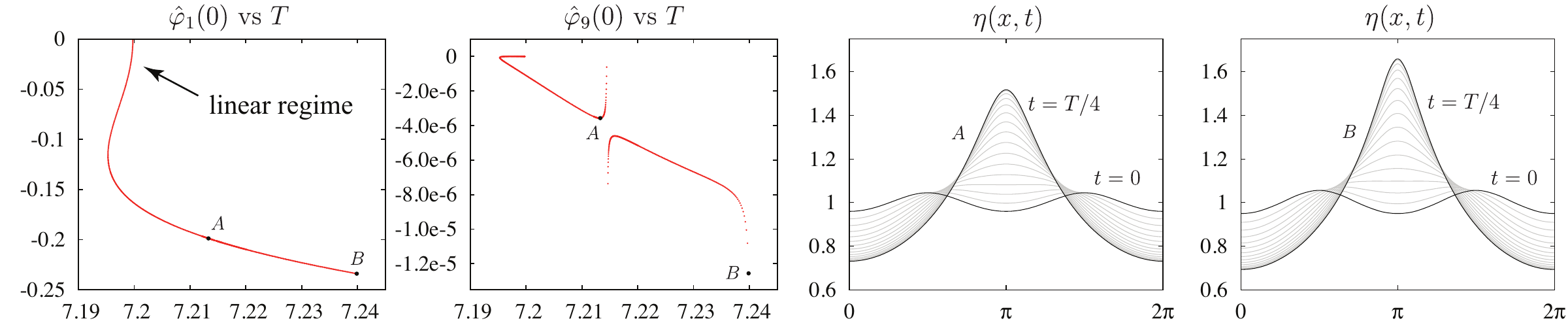}
\caption{\label{fig:bif10} A family of standing water waves of
  unit depth ($\hh=1$) bifurcates from the stationary solution at
  $T=2\pi/\sqrt{\tanh{1}}\approx7.200$. We used the ACM method to
  track the family out of the linearized regime via numerical
  continuation.  The period initially decreases with amplitude, but
  later increases to surpass the period of the linearized standing
  waves. A resonance near solution $A$ causes the 9th Fourier
  mode of $\varphi$ to jump discontinuously as the period
  increases. This resonance has little effect on the first Fourier
  mode.}
\end{figure}

To find time-periodic solutions of the nonlinear problem, we start
with a small amplitude linearized solution as an initial
guess. Holding $c_1=\hat\varphi_1(0)$ constant, we solve for the other
$c_k$ in (\ref{eq:init:stand}) using the ACM method of
Section~\ref{sec:acm}. Note that $c_1=B$ in the linearized regime.  We
then repeat this procedure for another value of $c_1$ to obtain a
second small-amplitude solution of the nonlinear problem.  The
particular choices we made were $c_1=-0.001$ and $c_1=-0.002$.  We then
varied $c_1$ in increments of $-0.001$, using linear extrapolation
from the previous two solutions for the initial guess.  The results
are shown in Fig.~\ref{fig:bif10}.  The two representative solutions
labeled A and B show that the amplitude of the wave increases and the
crest sharpens as the magnitude of $c_1$ increases. We chose $c_1$ to
be negative so the peak at $T/4$ would occur at $x=\pi$ rather than
$x=0$. An identical bifurcation curve (reflected about the $T$-axis)
would be obtained by increasing $c_1$ from 0 to positive values.  The
solutions A and B would then be shifted by $\pi$ in space.

%%%%%%%%%%%%%%%%%%%%%

For most values of $c_1$ between $0.0$ and $-0.23$, the ACM method has
no difficulty finding time-periodic solutions to an accuracy of
$f<10^{-26}$.  However, at $c_1=-0.201$, the minimum value of $f$
exceeds this target.  On further investigation, we found there was a
small gap, $c_1\in(-0.20113,-0.20124)$, where we were unable to
compute time-periodic solutions even after increasing $M$ from 256 to
512 and decreasing the continuation stepsize to $\Delta c_1 =
1.0\times 10^{-5}$.  By plotting other Fourier modes of the initial
conditions versus the period, we noticed that the 9th mode jumps
discontinuously when $c_1$ crosses this gap.  A similar disconnection
appears to be developing near solution B.

\begin{figure}
\includegraphics[width=\linewidth]{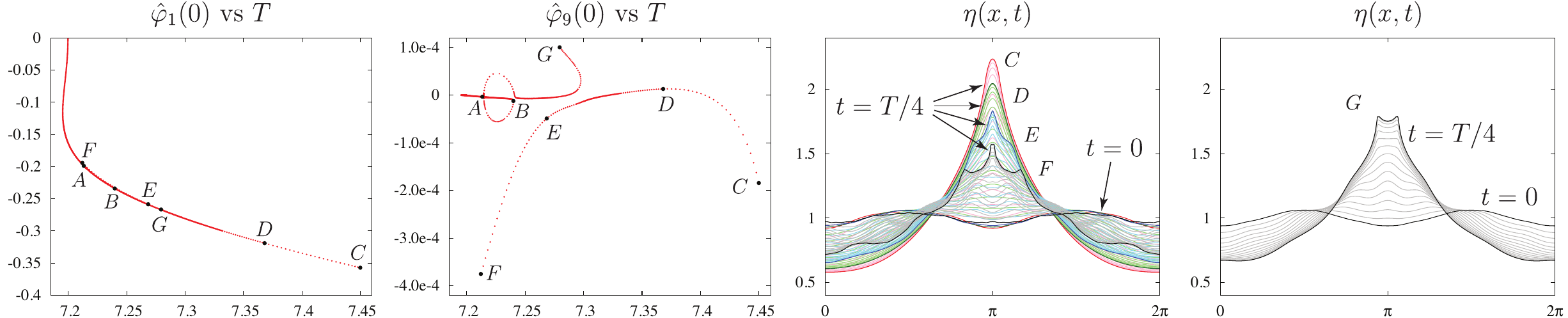}
\caption{\label{fig:bif100} Several branches of standing waves were
  found by extrapolation across disconnections in the bifurcation
  curves.  These disconnections are caused by resonant modes that may
  be interpreted physically as high-frequency standing waves
  superposed (nonlinearly) on the low-frequency carrier wave.
}
\end{figure}

Studying the results of Fig.~\ref{fig:bif10}, we suspected we could
find additional solutions by back-tracking from B to the region of the
bifurcation curve around $c_9=-7.0\times 10^{-6}$ and performing a
large extrapolation step to $c_9\approx-1.0\times 10^{-5}$, hoping to
jump over the disconnection at B.  This worked as expected, causing us
to land on the branch that terminates at G in Fig~\ref{fig:bif100}.
We used the same technique to jump from this branch to a solution
between E and D.  We were unable to find any new branches beyond C by
extrapolation from earlier consecutive pairs of solutions.

%%%%%%%%%%%%%%%%%%%%

Next we track each solution branch as far as possible in each
direction. This requires switching among the $c_k$ as bifurcation
parameters when traversing different regions of the solution space.
The period, $c_0=T$, is one of the options.  We also experimented with
pseudo-arclength continuation
\cite{keller:68,doedel91,smith:roberts:99}, but found that it is
necessary to re-scale the Fourier modes to successfully traverse folds
in the bifurcation diagram. This requires just as much human
intervention as switching among the $c_k$, so we abandoned the
approach.  The disconnections at A and B turn out to meet each other,
so that B is part of a closed loop and A is connected to the branch
containing G.  We stopped at G, F, C because the computations became
too expensive to continue further with the desired accuracy of
$f<10^{-26}$ using the adjoint continuation method.

The use of Fourier modes of the initial conditions in the bifurcation
diagrams is unconventional, but yields insight about the effect of
resonance on the dynamics of standing waves.  We observe
experimentally that disconnections in the bifurcation curves
correspond to higher-frequency standing waves appearing at the surface
of lower-frequency carrier waves. Because the equations are nonlinear,
only certain combinations of amplitude and phase can occur.
We generally see two possible solutions, one in which the high and
low-frequency component waves are in phase with each other, and
another where they are out of phase. For example, solutions F and G in
Fig.~\ref{fig:bif100} can both be described as a $k=7$ wave-number
standing wave oscillating on top of a $k=1$ carrier wave, but the
smaller wave sharpens the crest at F and flattens it at G, being 180
degrees out of phase at F versus G when the composite wave comes to
rest.  (All the standing waves of this paper reach a rest state at
$t=T/4$, by construction. Other types of solutions will be considered
in future work \cite{water:stable}.)  In Section~\ref{sec:bif:degen}, we show
that this disconnection between branches F and G is caused by a
$(3,7)$ harmonic resonance at fluid depth $h=1.0397$, where the period
of the $k=1$ mode is equal to 3 times the period of the $k=7$ mode for
small-amplitude waves \cite{smith:roberts:99, ioualalen:03}.

\begin{figure}
\includegraphics[width=\linewidth]{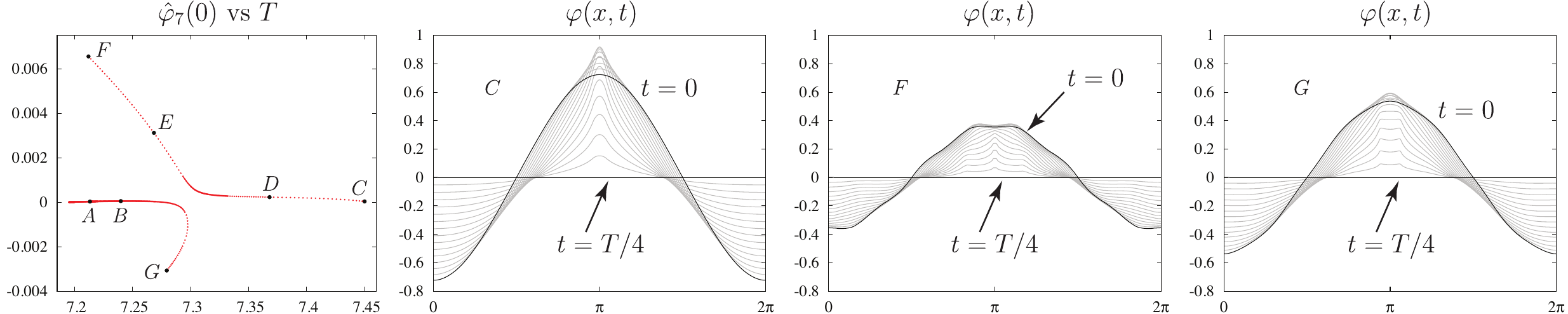}
\caption{\label{fig:bif100A}
  Bifurcation diagram showing $c_7=\hat\varphi_7(0)$ versus $T$ for
  standing waves of unit depth, along with snapshots of the evolution
  of $\varphi(x,t)$ for three of these solutions.  A secondary
  standing wave with wave number $k=7$ can be seen visibly superposed
  on $\varphi(x,0)$ in solution F, which corresponds to the large
  value of $c_7$ at F in the diagram.}
\end{figure}

In Fig.~\ref{fig:bif100A}, we plot $c_7=\hat\varphi_7(0)$ versus $T$,
along with the evolution of $\varphi(x,t)$ for several solutions over
time.  Note that the scale on the $y$-axis is 20 times larger here
(with $c_7$) than in Fig.~\ref{fig:bif100} (with $c_9$). This is why
the secondary standing waves in the plots of solutions F and G appear
to have wave number $k=7$. We also note that the disconnections at A
and B are nearly invisible in the plot of $c_7$ vs $T$.  This is
because the dominant wave number of these branches is $k=9$.
Similarly, it is difficult to observe any of the side branches in
the plot of $c_1$ vs $T$ in Fig.~\ref{fig:bif100} since they
all sweep back and forth along nearly the same curve. We will return
to this point in Section~\ref{sec:bif:degen}.

%%%%%%%%%%%%%%%%%%%%%

%%%%%%%%%%%%%%%%%%%%%
\begin{figure}
\includegraphics[width=\linewidth]{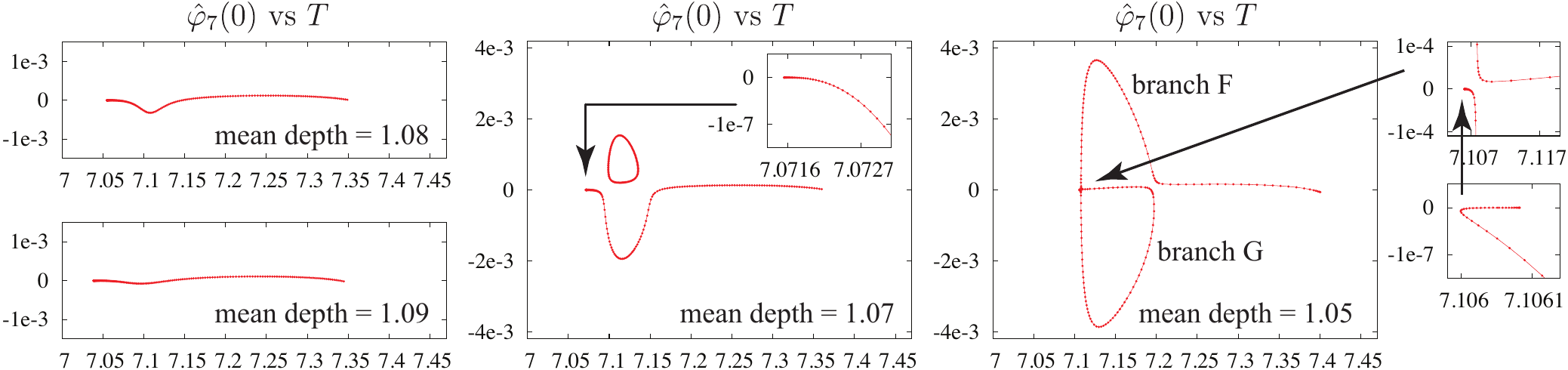}
\caption{\label{fig:bif105} If the mean depth, $h$, is increased from
  1.0 to 1.05, the loop structure between A and B in Fig.~\ref{fig:bif100}
  disappears, and branches F and G meet each other a second time at another
  imperfect bifurcation.
  As $h$ increases further, these loops shrink, disappearing
  completely by the time $h=1.09$.}
\end{figure}

\subsection{Nucleation of imperfect bifurcations}
\label{sec:nuc}

We next consider the effect of fluid depth on these bifurcation
curves. We found the ACM method was too slow to perform this study
effectively, which partly motivated us to develop the trust region
shooting algorithm.  As shown in Fig.~\ref{fig:bif105}, if the fluid
depth is increased from $h=1.0$ to $h=1.05$, it becomes possible to
track branches F and G to completion. The large amplitude oscillations
in the 7th Fourier mode eventually die back down when these branches
are followed past the folds at $c_7\approx \pm4\times 10^{-3}$ in
Fig.~\ref{fig:bif105}. The branches eventually meet each other at an
imperfect bifurcation close to the initial bifurcation from the
zero-amplitude state to the $k=1$ standing wave solutions.  This
imperfect bifurcation was not present at $h=1$.  Its nucleation will
be investigated in greater detail in Section~\ref{sec:bif:degen}.  The
small bifurcation loops at A and B in Fig.~\ref{fig:bif100} have
disappeared by the time $h=1.05$.  If we continue to increase $h$ to
$1.07$, the top wing of the S-shaped bifurcation loop breaks free from
the bottom wing and forms a closed loop.  This loop disappears by the
time $h$ reaches $1.08$.  By $h=1.09$, the $k=7$ resonance has all but
disappeared.

In Fig.~\ref{fig:bif10}, we saw that the period of standing waves of
unit depth decreases to a local minimum before increasing with wave
amplitude.  Two of the plots of Fig.~\ref{fig:bif105} show that this
remains true for $h=1.05$, but not for $h=1.07$. In the latter case,
the period begins increasing immediately rather than first decreasing
to a minimum.  This is consistent with the asymptotic analysis of
Tadjbakhsh and Keller \cite{tadjbakhsh}, which predicts that
\begin{equation}\label{eq:omega:correction}
  \jt
  \omega = \omega_0 + \frac{1}{2}\epsilon^2\omega_2 + O(\epsilon^3),
  \qquad \omega_0^2 = \tanh h, \qquad
  \omega_2 = \frac{1}{32}(9\omega_0^{-7} - 12\omega_0^{-3}
  -3 \omega_0 - 2\omega_0^5),
\end{equation}
where $\epsilon$ controls the wave amplitude, and agrees with
$A$ in (\ref{eq:lin:soln}) to linear order.  The correction term
$\omega_2$ is positive for $h<1.0581$ and negative for $h>1.0581$.

\begin{figure}
\includegraphics[width=\linewidth]{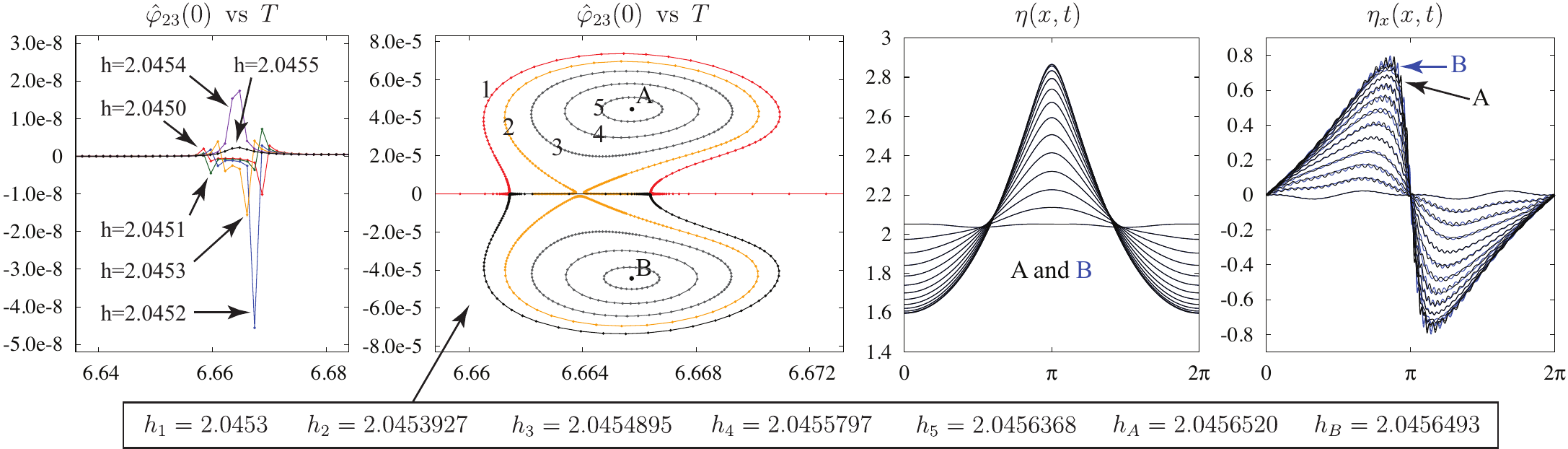}
\caption{\label{fig:bif205} A pair of imperfect bifurcations were
  found to coalesce as fluid depth increases, leaving behind two
  closed loops and a smooth bifurcation curve running between them.
  The loops each shrink to a point and disappear as fluid depth
  continues to increase.  Reversing the process shows that isolated
  solutions can nucleate new branches of solutions as fluid depth
  decreases. (right) The nucleated solutions A and B are nearly
  identical on large scales, but contain secondary, high-frequency
  standing waves at smaller scales that are out of phase with each
  other. These small oscillations become visible when the slope
  of the wave profile is plotted. }
\end{figure}

We will see in Section~\ref{sec:bif:degen} that the nucleation of
bifurcation branches between $h=1.09$ and $h=1.0$ is partly caused by
a $(3,7)$ harmonic resonance (defined below) at fluid depth
$h=1.0397$.  As this mechanism is complicated, we also looked for
simpler examples in deeper water.  The simplest case we found is shown
in Fig.~\ref{fig:bif205}.  For fluid depth $h=2$, we noticed a pair of
disconnections in the bifurcation curves that were not present for
$h=2.1$.  The 23rd Fourier mode of the initial condition exhibits the
largest deviation from 0 on the side branches of these disconnections.
However, as discussed in the next section, this is not caused by a
harmonic resonance of type $(m,23)$ for some integer~$m$.  To
investigate the formation of these side branches, we swept through the
region $6.64\le T\le 6.68$ with slightly different values of $h$,
using $c_0=T$ as the bifurcation parameter.  As shown in
Fig.~\ref{fig:bif205}, when $h=2.0455$, the bifurcation curve bulges
slightly but does not break.  As $h$ is decreased to $2.045$, a pair
of disconnections appear and spread apart from each other.  We
selected $h=2.0453$ as a good starting point to follow the side
branches. As we hoped would happen, the two red side branches in the
second panel of the figure met up with each other (at $c_{23}\approx
7\times 10^{-5}$), as did the two black branches (at $c_{23}\approx
-7\times 10^{-5}$).  We switched between $T$ and $c_{23}$ as
bifurcation parameters to follow these curves.  We then computed two
paths (not shown) in which $c_{23}=\pm 4\times 10^{-5}$ was held fixed
as $h$ was increased.  We selected 4 of these solutions to serve as
starting points to track the remaining curves in
Figure~\ref{fig:bif205}, which have fluid depths $h_2$ through $h_5$
given in the figure.  We adjusted $h_2$ to achieve a near three-way
bifurcation. This bifurcation is quite difficult to compute as the
Hessian of $f$ becomes nearly singular; for this reason, some of the
solutions had to be computed in quadruple precision to avoid falling
off the curves.  Finally, to find the points A and B where a single,
isolated solution exists at a critical depth, we computed $h$ as a
function of $(T,c_{23})$ on a small $10\times 10$ grid patch near A
and B, and maximized the polynomial interpolant using Mathematica.

%%%%%%%%%%%%%%%%%%%%%%%%

\subsection{Degenerate and secondary bifurcations due to harmonic
resonance}
\label{sec:bif:degen}

In this section, we explore the source of the resonance between the
$k=1$ and $k=7$ modes in water of depth $h$ close to 1.  While
harmonic resonances such as this have long been known to cause
imperfect bifurcations \cite{mercer:94, smith:roberts:99,
  ioualalen:03}, we are unaware that anyone has been able to track the
side-branches all the way back to the origin, where they meet up with
mixed-mode solutions of the type studied asymptotically by
Vanden-Broeck \cite{vandenBroeck:84} and numerically by Bryant and
Stiassnie \cite{bryant:stiassnie:94}.  In the traveling wave case,
such mixed-mode solutions are known as Wilton's ripples
\cite{chen:saffman:79, akers:wilton}.  When the fluid depth is
perturbed, we find that the degenerate bifurcation splits into a
primary bifurcation and two secondary bifurcations
\cite{bauer:keller}.  This is consistent with Bridges' work on
perturbation of degenerate bifurcations in three-dimensional standing
water waves in the weakly nonlinear regime \cite{bridges:standing:3d}.

We begin by observing that the ratio of the periods of two
small-amplitude standing waves is
\begin{equation}
  m = \frac{T_1}{T_2} = \frac{\omega_2}{\omega_1} =
  \sqrt{\frac{k_2\tanh k_2\hh}{k_1\tanh k_1\hh}}.
\end{equation}
If we require $m$ to be an integer and set $k_1=1$, we obtain
\begin{equation}\label{eq:harm:res}
  k_2\tanh k_2h = m^2\tanh h.
\end{equation}
Following \cite{mercer:94, smith:roberts:99, ioualalen:03}, we say
there is a harmonic resonance of order $(m,k_2)$ if $h$ satisfies
(\ref{eq:harm:res}).  At this depth, linearized standing waves of wave
number $k=1$ have a period exactly $m$ times larger than standing
waves of wave number $k=k_2$.  This nomenclature comes from the
short-crested waves literature \cite{ioualalen:93, ioualalen:96}; a
more general framework can be imagined in which $k_1$ is not assumed
equal to 1 and $m$ is allowed to be rational, but we do not need such
generality.  

We remark that the nucleation event discussed in the previous section
does not appear to be connected to a harmonic resonance.  In that
example, the fundamental mode must have a fairly large amplitude
before the secondary wave becomes active, and the secondary wave is
not a clean $k=23$ mode.  Also, no integer $m$ causes the fluid depth
of an $(m,23)$ resonance to be close to $2.045$.  The situation is
simply that at a certain amplitude, the $k=1$ standing wave excites a
higher-frequency, smaller amplitude standing wave that oscillates at
its surface.  It is not possible to decrease both of their amplitudes
to zero without destroying the resonant interaction in this case.

We now restrict attention to the $(3,7)$ harmonic resonance.  Setting
$m=3$ and $k_2=7$ in (\ref{eq:harm:res}) yields
\begin{equation}
  7\tanh 7h = 9\tanh h, \quad
  h>0 \qquad \Rightarrow \qquad
  h = h_\text{crit} \approx 1.0397189.
\end{equation}
In the nonlinear problem, when the fluid depth has this critical
value, we find that the $k=7$ and $k=1$ branches persist as if the
other were not present.  Indeed, the former can be computed as a
family of $k=1$ solutions on a fluid of depth $7h$.  The latter can be
computed by taking a pure $k=1$ solution of the linearized problem as
a starting guess and solving for the other Fourier modes of the
initial conditions, as before.  When this is done, after setting
$\epsilon = \hat\varphi_1(0)$, we find that
$\hat\varphi_3(0)=O(\epsilon^3)$, $\hat\varphi_5(0)=O(\epsilon^5)$,
$\hat\varphi_7(0)=O(\epsilon^5)$, and
$\hat\varphi_9(0)=O(\epsilon^7)$.  To obtain these numbers, we used 10
values of $\epsilon$ between $10^{-4}$ and $10^{-3}$ and computed the
slope of a log-log plot. The calculations were done in quadruple
precision with a 32 digit estimate of $h_\text{crit}$ to avoid
corruption by roundoff error.  If we repeat this procedure with
$h=1.0$, we find instead that $\hat\varphi_7(0)=O(\epsilon^7)$,
$\hat\varphi_9(0)=O(\epsilon^9)$.  Thus, the degeneracy of the
bifurcation at $h_\text{crit}$ appears to slow the decay rate of the
7th and higher modes, but not enough to affect the behavior at linear
order.

We were surprised to discover that two additional branches also
bifurcate from the stationary solution when $h=h_\text{crit}$.  For
these branches, we find that $\hat\varphi_k(0) = O(\epsilon^p)$, where
the first several values of $p$ are
\begin{equation*}
  \begin{array}{r|r}
    k & p \\ \hline
    1 & 1 \\
    3 & 3 \\
    5 & 3
    \end{array} \qquad\quad
  \begin{array}{r|r}
    k & p \\ \hline
    7 & 1 \\
    9 & 3 \\
    11 & 5
    \end{array} \qquad\quad
  \begin{array}{r|r}
    k & p \\ \hline
    13 & 3 \\
    15 & 3 \\
    17 & 5
    \end{array} \qquad\quad
  \begin{array}{r|r}
    k & p \\ \hline
    19 & 5 \\
    21 & 3 \\
    23 & 5
    \end{array} \qquad\quad
  \begin{array}{r|r}
    k & p \\ \hline
    25 & 7 \\
    27 & 5 \\
    29 & 5
    \end{array}
\end{equation*}
These numbers were computed as described above, with $\epsilon$
ranging between $10^{-4}$ and $10^{-3}$. To get a clean integer for
$\hat\varphi_{21}(0)$, $\hat\varphi_{27}(0)$ and
$\hat\varphi_{29}(0)$, we had to drop down to the range
$10^{-5}\le\veps\le10^{-4}$.  Using the Aitken-Neville algorithm
\cite{deuflhard}
to extrapolate $\hat\varphi_k(0)/\veps^p$ to $\veps=0$, we
obtain the leading coefficients for the two branches:
\begin{equation}
  \begin{array}{r|l}
     k & \qquad\qquad\hat\varphi_k(0) \\ \hline
     1 & \phm \epsilon \\
     7 & \phm 0.034152137008 \epsilon + O(\epsilon^3) \\
     3 &     -0.376330285335 \epsilon^3 + O(\epsilon^5) \\
     5 & \phm 0.065341882841 \epsilon^3 + O(\epsilon^5) \\
     9 &     -0.172818320378 \epsilon^3 + O(\epsilon^5) \\
    13 & \phm 0.019277463225 \epsilon^3 + O(\epsilon^5) \\
    15 &     -0.011062972892 \epsilon^3 + O(\epsilon^5) \\
    21 &     -1.303045\times 10^{-8} \epsilon^3 + O(\epsilon^5)
  \end{array}
  \qquad\qquad
  \begin{array}{r|l}
     k & \qquad\qquad\hat\varphi_k(0) \\ \hline
     1 & \phm \epsilon \\
     7 &     -0.034152137008 \epsilon + O(\epsilon^3) \\
     3 &     -0.376330285335 \epsilon^3 + O(\epsilon^5) \\
     5 &     -0.065341882841 \epsilon^3 + O(\epsilon^5) \\
     9 & \phm 0.172818320378 \epsilon^3 + O(\epsilon^5) \\
    13 & \phm 0.019277463225 \epsilon^3 + O(\epsilon^5) \\
    15 &     -0.011062972892 \epsilon^3 + O(\epsilon^5) \\
    21 & \phm 1.303045\times 10^{-8} \epsilon^3 + O(\epsilon^5)
  \end{array}
\end{equation}
In summary, there are four
families of solutions of the nonlinear problem that bifurcate from the
stationary solution. In the small amplitude limit, they approach a
pure $k=1$ mode, a pure $k=7$ mode, and two mixed modes involving both
$k=1$ and $k=7$ wave numbers. For convenience, we will refer to these
branches as ``pure'' and ``mixed'' based on their limiting behavior
in the linearized regime. The mixed mode solutions are examples
of the Wilton's ripple phenomenon \cite{vandenBroeck:84,
  chen:saffman:79, akers:wilton} in which multiple wavelengths
are present in the leading order asymptotics.

%%%%%%%%%%%%%%%%%%%%%%
\begin{figure}
\includegraphics[width=\linewidth]{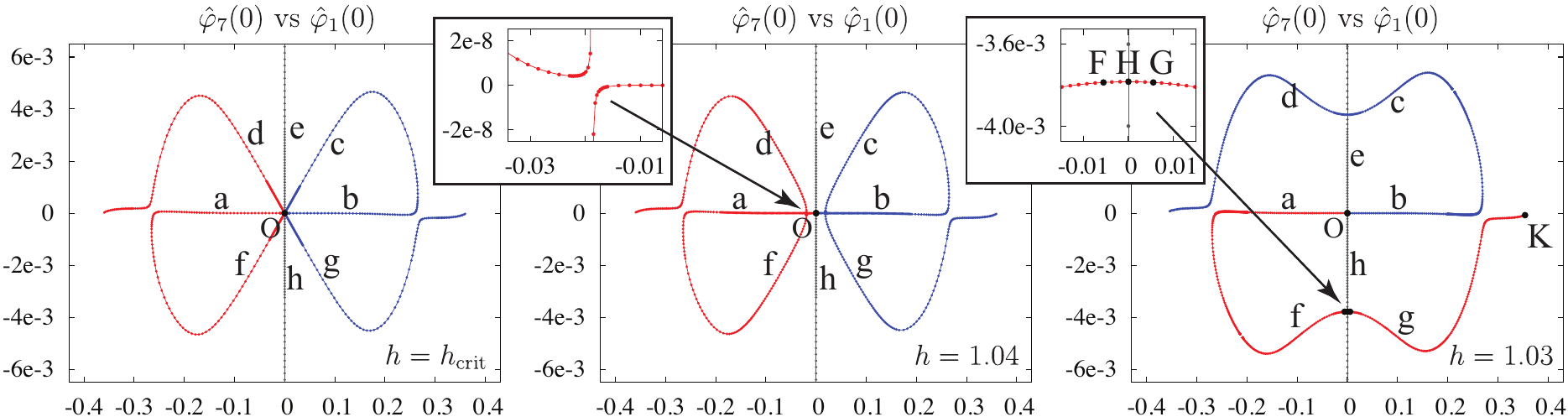}
\caption{\label{fig:spider} Perturbation of this degenerate
  bifurcation causes a pair of imperfect ($h>h_\text{crit}$) or
  perfect ($h<h_\text{crit}$) secondary bifurcations to form.  Red
  markers are the solutions actually computed, while blue markers
  correspond to the same solutions, phase-shifted in space by $\pi$.
}
\end{figure}

When these four branches are tracked in both directions, we end up
with eight rays of solutions emanating from the equilibrium
configuration, labeled a--h in Figure~\ref{fig:spider}. Rays a and b
consist of pure $k=1$ mode solutions, with negative and positive
amplitude, respectively, where amplitude refers to $\hat\varphi_1(0)$.
Rays e and h are the pure $k=7$ mode solutions, and rays c,d,f,g are
the mixed mode solutions.  It is remarkable that rays a and f, as well
as b and c, are globally connected to each other by a large loop in
the bifurcation diagram.  By contrast, for the Benjamin-Ono equation
\cite{benj3}, additional branches of solutions that emanate from a
degenerate bifurcation belong to different levels of the hierarchy of
time-periodic solutions than the main branches; thus, solutions on the
additional branches have a different number of phase parameters, and
cannot meet up with one of the main branches without another
bifurcation.

We now investigate what happens to these rays when the fluid depth is
perturbed.  When $h$ increases from $h_\text{crit}$ to $1.04$, rays e
and h (the pure $k=7$ solutions) break free from the other 6 rays.  An
imperfect bifurcation forms on rays a and b, linking the former to f
and d, and the latter to c and g. Aside from this local reshuffling of
branch connections near the stationary solution, the global
bifurcation structure of $h=1.04$ is similar to $h=h_\text{crit}$.  In
the other direction, when $h=1.03<h_\text{crit}$, rays a and b (the
pure $k=1$ solutions) disconnect from the other rays.  Instead of
forming imperfect bifurcations as before, rays c and d separate from
the $\hat\varphi_7=0$ axis, but remain connected to ray e through a
perfect bifurcation.  The same is true of rays f, g and h. Thus, we
have identified a case where perturbing a degenerate bifurcation
causes it to break up into a primary bifurcation and two secondary
bifurcations \cite{bauer:keller}, either perfect ($h<h_\text{crit}$)
or imperfect ($h>h_\text{crit}$).

The reason one is perfect and the other is not can be explained
heuristically as follows.  All the solutions on the pure $k=7$ branch
have Fourier modes $\hat\varphi_k(t)$, with $k$ not divisible by 7,
exactly equal to zero.  These modes can be eliminated from the
nonlinear system of equations by reformulating the problem as a $k=1$
solution on a fluid of depth $7h$. This reformulation removes the
resonant interaction by restricting the $k=1$ mode (in the original
formulation) to remain zero. The simplest way for this mode to become
non-zero, i.e.~deviate from rays e,h in Fig.~\ref{fig:spider}, is
through a subharmonic bifurcation (with $k=7$ as the fundamental
wavelength) in which the Jacobian $J$ in (\ref{eq:Jik}) develops a
non-trivial kernel containing a null vector $c\in l^2(\mathbb{N})$
with $c_1=\hat\varphi_1(0)\ne0$. Here $c$ contains the even modes of
$\eta$ and the odd modes of $\varphi$ at $t=0$, as in
(\ref{eq:init:stand}), but with $n=\infty$.  If such a kernel exists,
one expects to be able to perturb the solution in this direction,
positively or negatively, to obtain a pitchfork bifurcation.  By
contrast, solutions on the $k=1$ branch develop non-zero
higher-frequency modes through non-linear mode interactions.  So while
$c_1=0$ on the $k=7$ branch, $c_7\ne0$ on the $k=1$ branch.  Since
there is no way to control the influence of the 7th mode, e.g.~by
constraining it to be zero, there really is no ``pure'' $k=1$ branch
to bifurcate from, and the result is an imperfect bifurcation.

%%%%%%%%%%%%%%%%%%%%%%
\begin{figure}
\includegraphics[width=\linewidth]{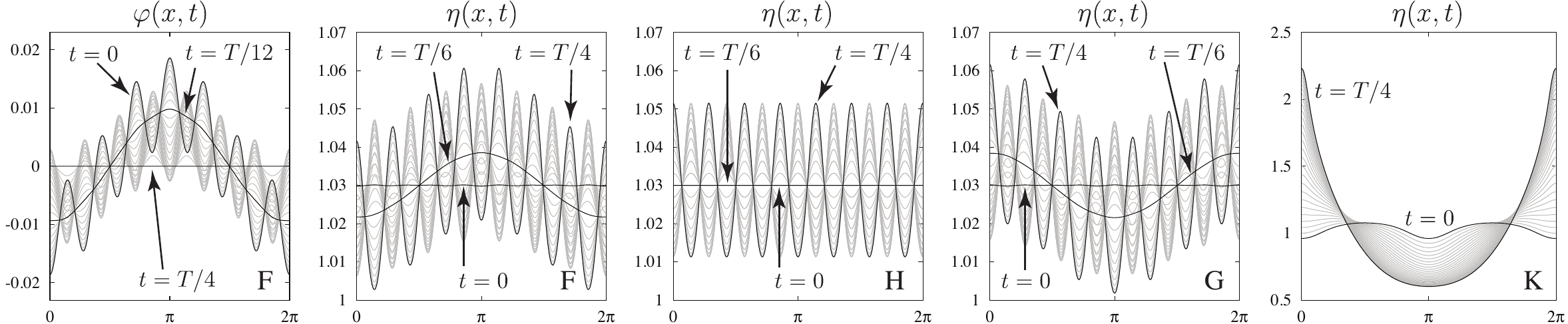}
\caption{\label{fig:104plots} Solutions labeled F, G, H and K in
  Fig.~\ref{fig:spider} show the transition from standing waves that
  form crests at $x=\pi$ to those that form crests at $x=0$ when
  $t=T/4$. This transition occurs where branch f meets branch g in
  the $h<h_\text{crit}$ case.  For $h>h_\text{crit}$, branch f meets
  branch d at an imperfect bifurcation, and the solution at the
  end of path d would resemble solution K,
  shifted in space by $\pi$.
}
\end{figure}

The fact that ray f is connected to g for $h<h_\text{crit}$, and to d
for $h>h_\text{crit}$, has a curious effect on the form of the
numerical solution at the end of the red branch, the branch of
solutions actually computed, in Fig.~\ref{fig:spider}. In the former
case, $c_1$ changes sign from branch f to g, and we end up at solution
K in Fig.~\ref{fig:104plots}, which forms a wave crest at $x=0$ at
$t=T/4$.  In the latter case, $c_1$ remains negative from branch f to
d (or branch a to d if the imperfect bifurcation is traversed without
branch jumping) and we end up at a solution similar to K, but phase
shifted, so that a wave crest forms at $x=\pi$ at $t=T/4$.  Note that
the sign of $c_1$ determines whether the fluid starts out flowing
toward $x=\pi$ and away from $x=0$, or vice-versa.

The transition from wave crests at $x=\pi$ to wave crests at $x=0$
when $t=T/4$ is shown in Fig.~\ref{fig:104plots}. Solutions F, G and H
may all be described as $k=7$ standing waves superposed on $k=1$
standing waves. Note that solution F bulges upward at $x=\pi$ when
$t=T/4$, while solution G bulges downward there.  A striking feature
of these plots is that the $k=7$ modes of $\varphi$ and $\eta$ nearly
vanish at $t=T/12$ and $t=T/6$, respectively.  This occurs because the
$k=7$ mode oscillates 3 times faster than the $k=1$ mode. Solution H
is a pure $k=7$ solution, which means $\varphi(x,t)$ vanishes
identically at $t=\frac{2m+1}{4}\left(\frac{T}{3}\right)$, $m\ge0$,
while $\hat\eta_7(t)$ passes through zero at
$t=\frac{m}{2}\left(\frac{T}{3}\right)$, $m\ge0$.  Since solutions F
and G are close to solution H, $\hat\varphi_7(t)$ and $\hat\eta_7(t)$
pass close to zero at these times, leading to smoother solutions
dominated by the first Fourier mode at these times.

%%%%%%%%%%%%%%%%%%%%%%
\begin{figure}
\includegraphics[width=\linewidth]{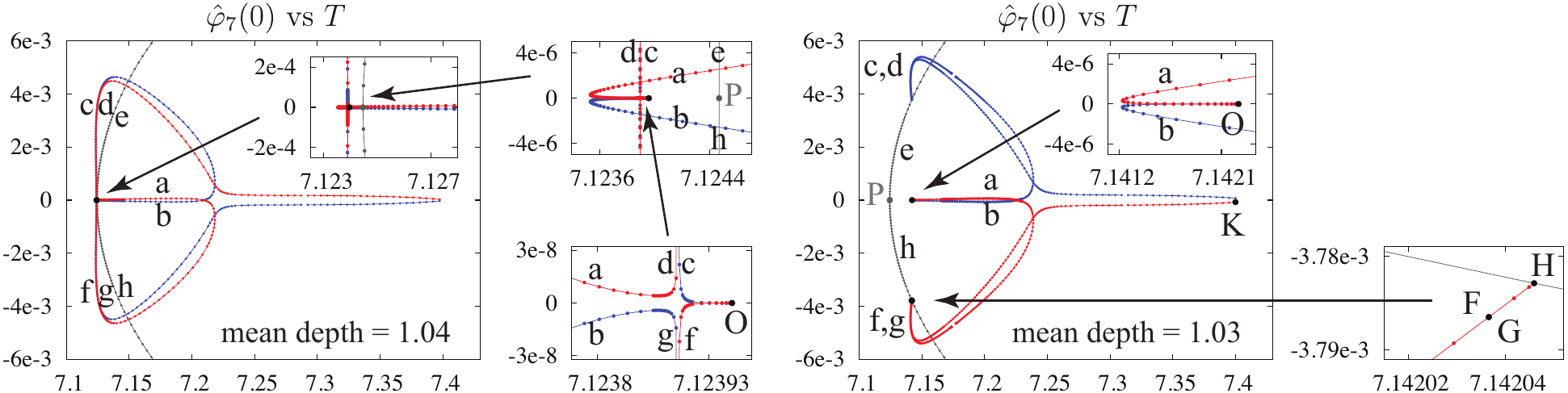}
\caption{\label{fig:bif104} 
  % If we plot the period on the $x$-axis
  Solutions O and P, the bifurcation points from the equilibrium state
  to the pure $k=1$ and $k=7$ standing waves, respectively, separate
  from each other through a change in period as fluid depth varies
  from $h=h_\text{crit}$. The imperfect bifurcation can occur arbitrarily
  close to solution O by taking $h\searrow h_\text{crit}$.
  Turning points in $\hat\varphi_7(0)$ and $T$ occur at solution H
  for $h<h_\text{crit}$.
}
\end{figure}

Figure~\ref{fig:bif104} shows how the period varies along each of
these solution branches. The period varies with fluid depth more
rapidly for solutions on the $k=1$ branch than on the $k=7$ branch
since the slope of $\tanh h$ is 18000 times larger than that of $\tanh
7h$ when $h\approx 1$.  As a result, bifurcation point O (to the $k=1$
branch) moves visibly when $h$ changes from $1.03$ to $1.04$, while
bifurcation point P (to the $k=7$ branch) hardly moves at all.  We
also see that the period increases with amplitude on branches e and h.
By contrast, on branches a and b, it decreases to a local minimum
before increasing with amplitude.  This is consistent with the
asymptotic analysis of Tadjbakhsh and Keller discussed previously; see
(\ref{eq:omega:correction}) above.  Finally, we note that both $T$ and
$c_7$ have a turning point at solution H, the bifurcation point
connecting branches f and g to branch h.  This causes paths f and g to
lie nearly on top of each other for much of the bifurcation
diagram. Other examples of distinct bifurcation curves tracing back
and forth over nearly the same paths are present (but difficult to
discern) in Figures~\ref{fig:spider} and~\ref{fig:bif104} when
$h=1.03$, and will be discussed further in Section~\ref{sec:conclude}.

\subsection{Breakdown of self-similarity and the Penney and Price conjecture}
\label{sec:breakdown}

In Sections~\ref{sec:nuc} and~\ref{sec:bif:degen} above, we have seen
that increasing the fluid depth causes disconnections in the
bifurcation diagrams to ``heal,'' and it is natural to ask if any will
persist to the infinite depth limit. The answer turns out to be yes,
which is not surprising from a theoretical point of view since
infinite depth standing waves are completely resonant \cite{iooss05},
involving state transition operators with infinite dimensional kernels
and a small-divisor problem on the complement of this kernel.
Nevertheless, examples of such disconnecitons have only recently been
observed in numerical simulations \cite{breakdown}, and show no
evidence of being densely distributed along bifurcation curves.  In
this section, we expand on the results of \cite{breakdown}, filling in
essential details and providing new material not discussed there. The
scarcity of observable disconnections will be discussed further in the
conclusion section.

The main question addressed in \cite{breakdown} is whether standing
waves of extreme form approach a limiting wave profile with a
geometric singularity at the wave crest when the bifurcation curve
terminates. This type of question has a long history, starting with
Stokes \cite{stokes:1880,craik:05}, who predicted that the periodic
traveling wave of greatest height would feature wave crests with
sharp, $120^\circ$ interior crest angles.  While there are some
surprises concerning oscillatory asymptotic behavior at the crest of
the almost highest traveling wave \cite{lhf:77, lhf:78, breakdown}, it
has been confirmed both theoretically \cite{amick:82} and numerically
\cite{maklakov,gandzha:07} that a limiting extreme traveling wave does
exist, and possesses a sharp $120^\circ$ wave crest.  For standing
waves, a similar conjecture was made by Penney and Price in 1952
\cite{penney:52}, who predicted that the limiting extreme wave would
develop sharp, 90 degree interior crest angles each time the fluid
comes to rest. As discussed in the introduction, numerous
experimental, theoretical and numerical studies \cite{taylor:53,
  grant, okamura:98, okamura:03, okamura:10, mercer:92, schultz} have
reached contradictory conclusions on the limiting behavior at the
crests of extreme standing waves.

\begin{figure}
\includegraphics[width=\linewidth]{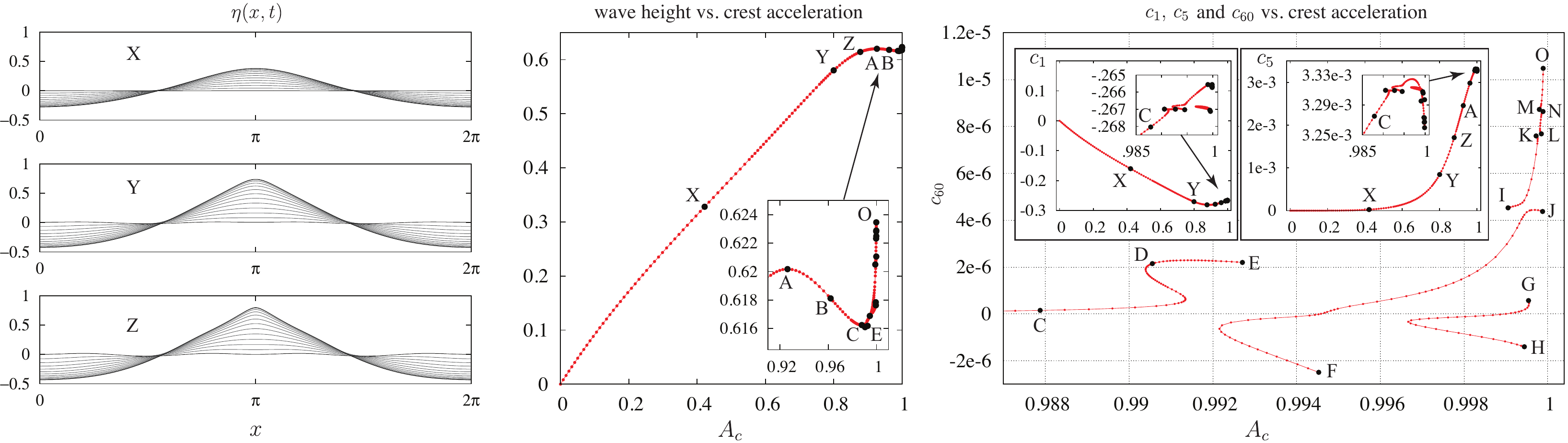}
\caption{\label{fig:bif:00} Snapshots of several standing waves over a
  quarter period in water of infinite depth, along with bifurcation
  diagrams showing where they fit in. (center) Conventional
  bifurcation diagram showing wave height versus crest
  acceleration. The turning point at A was discovered by Mercer and
  Roberts \cite{mercer:92} while the turning point at C and subsequent
  bifurcation structure were discovered by Wilkening \cite{breakdown}.
  The wave height $h$ eventually exceeds the local maximum at A.
  (right) The continuation
  parameters actually used were $c_1$, $c_5$, $c_{60}$ and $T$ rather
  than $h$ or $A_c$. }
\end{figure}

\begin{figure}
\includegraphics[width=\linewidth]{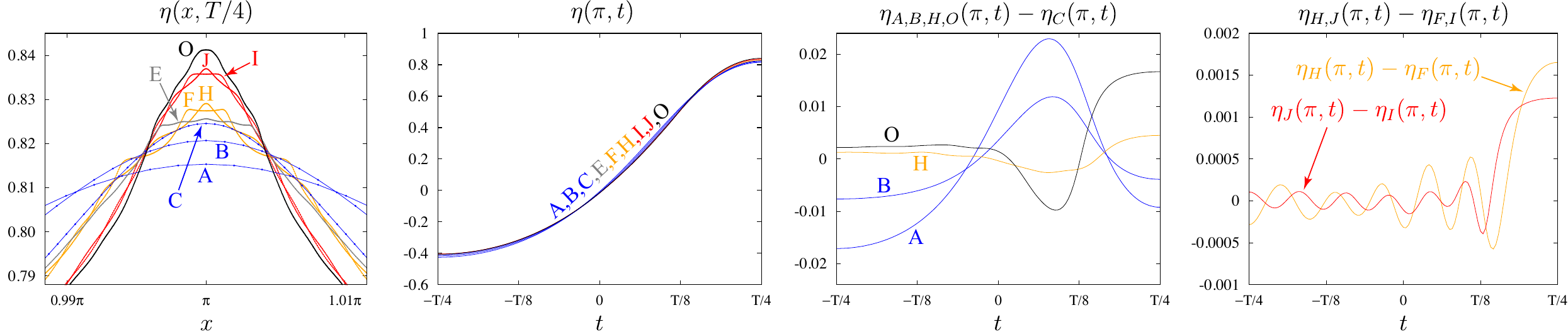}
\caption{\label{fig:tip} Oscillatory structures near the wave crest, and
  time-evolution of a particle from trough to crest, for several
  extreme standing waves. Labels correspond to the bifurcation
  diagrams in Figure~\ref{fig:bif:00}. These solutions take the form
  of higher-frequency standing waves superimposed nonlinearly on
  lower-frequency carrier waves. The higher-frequency oscillations
  occur both in space and time.  }
\end{figure}

Penney and Price expected wave height, defined as half the maximum
crest-to-trough height, to increase monotonically from the
zero-amplitude equilibrium wave to the extreme wave.  Mercer and
Roberts found that wave height reaches a turning point, achieving a
local maximum of $h=0.62017$ at $A_c=0.92631$, where $A_c$ is the
downward acceleration of a fluid particle at the wave crest at the
instant the fluid comes to rest (assuming $g=1$ in
(\ref{eq:ww})). Since $h$ is not a monotonic function, they proposed
using crest acceleration as a bifurcation parameter instead. However,
as shown in Fig.~\ref{fig:bif:00}, crest acceleration also fails to be
a monotonic function.
The bifurcation curve that was supposed to terminate at the extreme
wave when $A_c$ reaches 1 becomes fragmented for $0.99<A_c<1$.  Just
as in the finite depth case, this fragmentation is due to resonant
interactions between the large-scale carrier wave and various
smaller-scale, secondary standing waves that appear at the
surface of the primary wave.  Figure~\ref{fig:tip} shows several
examples of the oscillatory structures that are excited by resonance,
both in space and in time.  
The amplitude of the higher-frequency oscillations are small enough in
each case that the vertical position of a particle traveling from
trough to crest increases monotonically in time; however, plotting
differences of solutions on nearby branches as a function of time
reveals the temporal behavior of the secondary standing waves. We note
that the frequency of oscillation of the secondary waves (near
$x=\pi$) decreases as $t$ approaches $T/4$.
This seems reasonable as fluid particles at the crest are nearly in
free-fall at this time when $A_c$ is close to 1; thus, the driving
force of the secondary oscillations is low there.  In shallow water,
the secondary oscillations are often strong enough to lead to
non-monotonic particle trajectories from trough to crest (see
Section~\ref{sec:shallow}).

\begin{figure}
\includegraphics[width=\linewidth]{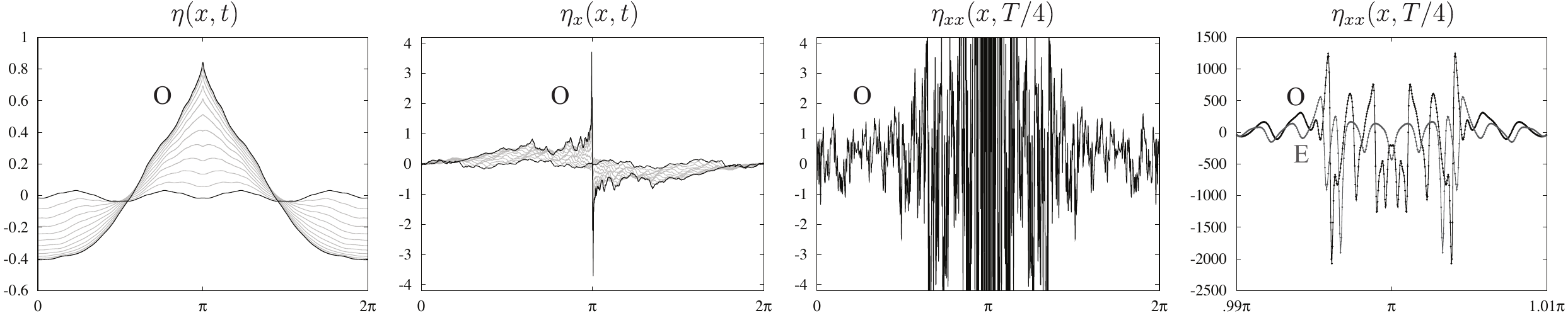}
\caption{\label{fig:breakdown} Evolution of surface height and its
  derivatives over a quarter period for solution O in
  Figure~\ref{fig:bif:00}. The plots at right of $\eta_{xx}$ are shown
  only at time $T/4$ for clarity. In the far right panel, we also
  plotted solution E for comparison. Each solution that terminates
  a branch in the bifurcation diagram is highly oscillatory;
  we followed the branches to the point that the computations became
  too expensive to continue further.
}
\end{figure}

If a limiting wave profile does not materialize as $A_c$ approaches 1,
a natural question arises as to what will terminate the bifurcation
curves.
In \cite{breakdown}, it was emphasized that oscillations at the crest
tip prevent self-similar sharpening to a corner, as happens in the
traveling case.  We note here that the entire wave
profile, not just the crest tip, develops high-frequency oscillations
on small scales toward the end of each bifurcation curve. This is
illustrated in Fig.~\ref{fig:breakdown}, and suggests that if these
bifurcation curves do end somewhere, without looping back to merge
with another disconnection, it may be due to solutions becoming
increasingly rough, with some Sobolev norm diverging in the limit.
We also remark that since many of these standing waves come close to
forming a $90^\circ$ corner, there may well exist nearly time-periodic
solutions that do pass through a singular state.  Taylor's thought
experiment \cite{taylor:53} in which water is piled up in a crested
configuration and released from rest could be applied to a sharply
crested perturbation of the rest state of one of our standing waves.
However, like Taylor, we see no reason that $90^\circ$ would be the
only allowable crest angle. It is conceivable that $90^\circ$ is the
only angle for which smooth solutions can propagate forward from a
singular initial condition, but we know of no such results.

\begin{figure}
\includegraphics[width=\linewidth]{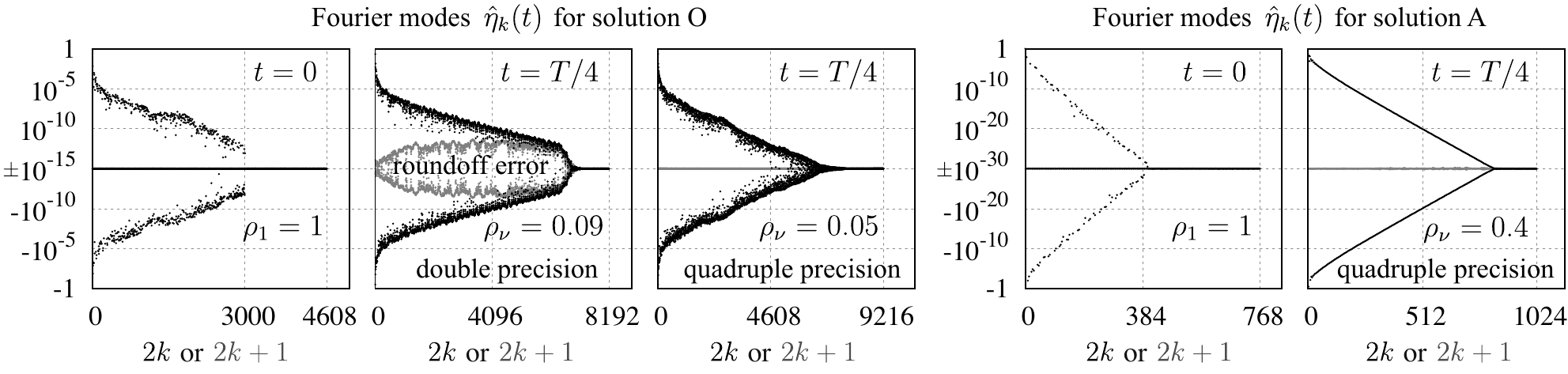}
\caption{\label{fig:fourier} The Fourier modes of $\eta$ (shown) and
  $\varphi$ (not shown) are monitored to decide how many grid points
  are needed to resolve the solution. The parameter $\rho_l$ controls the
  nonuniform spacing of gridpoints via (\ref{eq:xi}). The real (black)
  and imaginary (grey) parts of $\hat{\eta}_k(t)$ are plotted in
  positions $2k$ and $2k+1$, respectively.  (Left) the minimization
  was performed in double precision to obtain these initial
  conditions.  The result was checked in quadruple precision to
  eliminate roundoff error.  (Right) the minimization was performed in
  quadruple precision, yielding $f=2.1\times10^{-60}$.  }
\end{figure}

The increase in roughness of the solutions as crest acceleration
approaches $1$ may also be observed by plotting Fourier mode
amplitudes for various solutions along the bifurcation curve.  In
Fig.~\ref{fig:fourier}, we compare the Fourier spectrum of $\eta$ at
$t=0$ and $t=T/4$ for solutions O and A.  In both of these
simulations, we parametrized the curve non-uniformly, as in
(\ref{eq:xi}), to increase resolution near the crest tip.  Thus, a
distinction must be made between computing Fourier modes with respect
to $x$ versus $\alpha$:
\begin{equation}\label{eq:fourier}
  \hat\eta_k(t) = \frac{1}{2\pi}\int_0^{2\pi} \eta(x,t)e^{-ikx}\,dx, \qquad
  \text{ or } \qquad
  \hat\eta_k(t) = \frac{1}{2\pi}\int_0^{2\pi}
  \eta(\xi_{l(t)}(\alpha),t)e^{-ik\alpha}\,d\alpha.
\end{equation}
In this figure, we use the latter convention, since $\eta\circ\xi_l$
and $\varphi\circ\xi_l$ are the quantities actually evolved in time,
and the decay rate of Fourier modes is faster with respect to
$\alpha$.  At $t=0$, the two formulas in (\ref{eq:fourier}) agree
since we require $\xi_1(\alpha)=\alpha$.  For solution O, the Fourier
modes of the initial conditions decay to $|c_k|<10^{-12}$ for
$2k\ge3000$.  At $t=T/4$ we have
$\max(|\hat\eta_k|,|\hat\varphi_k|)<10^{-12}$ for $2k\ge6500$ with
$\rho_\nu=0.09$. For solution A, we have $|c_k|<10^{-29}$ for $2k\ge
400$ and $\max(|\hat\eta_k|,|\hat\varphi_k|)<10^{-29}$ for $2k\ge 800$
at $t=T/4$ with $\rho_\nu=0.4$. Here $T=1.629324$ for solution O and
$T=1.634989$ for solution A.  The mesh parameters used in these
simulations are listed in Table~\ref{tab:param:OA}.  We remark that
for solutions such as O with fairly sharp wave crests, decreasing
$\rho_\nu$ generally leads to faster decay of Fourier modes, but also
amplifies roundoff errors due to closer grid spacing near the
crest. Further decrease of $\rho_\nu$ in the double-precision
calculation does more harm than good.

\begin{table}[h]
\begin{equation*}
  \begin{array}{l||c||c|c|c|c||c|c|c|c||c||c|c|c|c||c|c|c|c}
    \text{\;\;solution} & \nu & \theta_1 & \theta_2 & \theta_3 & \theta_4 & \rho_1 & \rho_2 & \rho_3 & \rho_4 & n & M_1 & M_2 & M_3 & M_4 &  N_1 & N_2 & N_3 & N_4 \\ \hline
    \text{A (quad)} & \js 2 & \js 0.2 & \js 0.8 & \js - & \js - & \js 1.0 & \js 0.4 & \js - & \js - & \js 200 & \js 768 & \js 1024 & \js - & \js - & \js 24 & \js 144 & \js - & \js - \\
    \text{O (double)} & \js 4 & \js 0.2 & \js 0.3 & \js 0.3 & \js 0.2 & \js 1.0 & \js 0.4 & \js 0.1 & \js 0.09 & \js 1500 & \js 4608 & \js 6144 & \js 6912 & \js 8192 & \js 192 & \js 432 & \js 576 & \js 480 \\
    \text{O (quad)} & \js 4 & \js 0.1 & \js 0.3 & \js 0.4 & \js 0.2 & \js 1.0 & \js 0.25 & \js 0.08 & \js 0.05 & \js 1500 & \js 6144 & \js 7500 & \js 8192 & \js 9216 & \js  60 & \js 216 & \js 384 & \js 240
  \end{array}
\end{equation*}
\caption{\label{tab:param:OA}
Mesh parameters used to compute solutions A and O in Figures~\ref{fig:fourier}
and~\ref{fig:evol:PhiOA}.
}
\end{table}

The effect of roundoff-error and the 36th order filter can both be
seen in the second panel of Fig.~\ref{fig:fourier}. In exact
arithmetic, $\eta(x,t)$ and $\varphi(x,t)$ would remain even functions
for all time. However, in numerical simulations, the imaginary parts
of $\hat\eta_k(t)$ and $\hat\varphi_k(t)$ drift away from zero, giving
a useful indicator of how much the solution has been corrupted by
roundoff error. The filter (\ref{eq:filter}) has little effect on the
first 70 percent of the Fourier modes, but strongly damps out the last
15 percent.  By monitoring the decay of Fourier modes through plots
like this, one can ensure that the simulations are fully resolved, and
that filtering does not introduce more error than is already
introduced by roundoff error. We also monitor energy conservation,
\begin{equation*}
  E(t) = \frac{1}{2} \int_0^{2\pi} \left[\varphi(x,t)\mc{G}\varphi(x,t) +
  g\eta(x,t)^2\right]\,dx,
\end{equation*}
choosing time-steps small enough that $E$ remains constant to as many
digits as possible, typically 14 in double-precision and 29 in
quadruple precision. Note that $\mc{G}$ depends on time through
$\eta$.

\begin{figure}
\begin{center}
  \includegraphics[width=\linewidth]{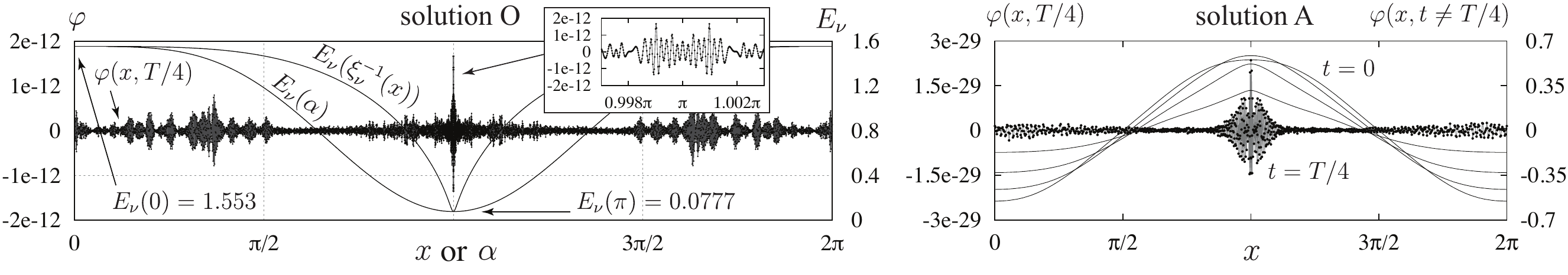}
\end{center}
\caption{\label{fig:evol:PhiOA}
  Plots of the residual $\varphi(x,T/4)$ for solutions O and A.
  (left) When minimized in double-precision arithmetic, we obtain
  $f=1.3\times 10^{-26}$.  Shown here is a re-computation of the
  solution in quadruple precision on a finer mesh using the same
  initial conditions. This yields $f=8.6\times 10^{-27}$, which is
  even smaller than predicted in double-precision. Re-spacing the grid
  maps the curve $E_\nu(\alpha)$ to $E_\nu(\xi_\nu^{-1}(x))$,
  improving the resolution that can be achieved with 9216 gridpoints.
  Note that the oscillations in $\varphi(x,T/4)$ are fully resolved.
  (right) The minimization was performed in quadruple-precision
  arithmetic, yielding $f=2.1\times 10^{-60}$. The velocity potential
  is nearly 30 orders of magnitude smaller at $t=T/4$ than at $t=0$.
}
\end{figure}

Because solution A remains smoother and involves many fewer Fourier
modes than solution O, it was possible for us to perform the entire
computation in quadruple precision arithmetic. This allowed us to
reduce $f$ in (\ref{eq:f:phi}) to $2.1\times 10^{-60}$. As shown in
Fig.~\ref{fig:evol:PhiOA}, the velocity potential of this solution
drops from $O(1)$ at $t=0$ to less than $3\times 10^{-29}$ at $t=T/4$,
in the uniform norm.  While it was not possible to perform the
minimization for solution O in quadruple precision arithmetic (due to
memory limitations of the GPU device), we were able to check the
double-precision result in quadruple precision to confirm that $f$ is
not under-predicted by the minimization procedure.
Because the DOPRI8 and SDC15 methods involve 12 and 99 internal
Runge-Kutta stages per time-step, respectively, more function
evaluations were involved in advancing the quadruple-precision
calculations through time even though $N_l$ is larger in the
double-precision calculations. As shown in Fig.~\ref{fig:evol:PhiOA},
the oscillations in $\varphi(x,T/4)$ remain fully resolved in the
quadruple precision calculation; thus, $f=8.6\times 10^{-27}$ is an
accurate measure of the squared error.  The predicted value of $f$ in
double-precision (obtained by minimizing $f$) was $f=1.3\times
10^{-26}$. Since minimizing $f$ entails eliminating as many
significant digits of $\varphi(x,T/4)$ as possible, the resulting
value of $f$ is not expected to be highly accurate. What is important
is that minimizing $f$ in double-precision does not grossly
underestimate its minimum value.  In fact, its value is often
over-estimated, as occurred here.  This robustness is a major benefit
of posing the problem as an \emph{overdetermined} non-linear
system. The only way to achieve a small value of $f$ is to accurately
track a solution of the PDE for which the exact $f$ is small.
Roundoff errors and truncation errors will cause the components of $r$
in (\ref{eq:f:phi}) to drift away from zero, leading to an
incompatible system of equations with minimum residual of the order of
the accumulated errors.

There is a big advantage to choosing $t=0$ to occur at the midpoint
between rest states rather than at a rest state.  The reason is that
many more Fourier modes are needed to represent $\eta$ when the wave
crest is relatively sharp, which for us occurs when $t$ is near
$T/4$. For example, solutions O and A in Fig.~\ref{fig:fourier} have
more than twice as many active Fourier modes at $t=T/4$ as they did at
$t=0$, even using a non-uniform grid to better resolve the crested
region.  Setting up the problem this way leads to fewer Fourier modes
of the initial condition to solve for, and increases the number of
non-linear equations. Thus, the system is more overdetermined, adding
robustness to the computation.

We conclude this section by mentioning that ``branch jumping'' is very
easy to accomplish (and hard to avoid) in the numerical continuation
algorithm. For strong disconnections such as at solution B in
Fig.~\ref{fig:bif10}, it is sometimes necessary to backtrack away from
the disconnection and then take a big step, hoping to land beyond the
gap. Since we measure the residual error in an overdetermined fashion,
it is obvious if we land in a gap where there is no time-periodic
solution --- the minimum value of $f$ remains large in that case.
However, most disconnections can be traversed without backtracking, or
even knowing in advance of their presence.  The disconnections in
Figure~\ref{fig:bif:00} were all discovered by accident in this way.
Once a disconnection is observed, we can go back and follow side
branches to look for global re-connections or new families of
time-periodic solutions.

\subsection{Counter-propagating solitary waves in shallow water}
\label{sec:shallow}

In previous sections, we saw that decreasing the fluid depth causes
nucleation of loop structures in the bifurcation curves that nearly
(or actually) meet at imperfect (or perfect) bifurcations.  Some of
these disconnections persist in the infinite depth limit.  We now
consider the other extreme of standing waves in very shallow water.

\begin{figure}
\begin{center}
\includegraphics[width=\linewidth]{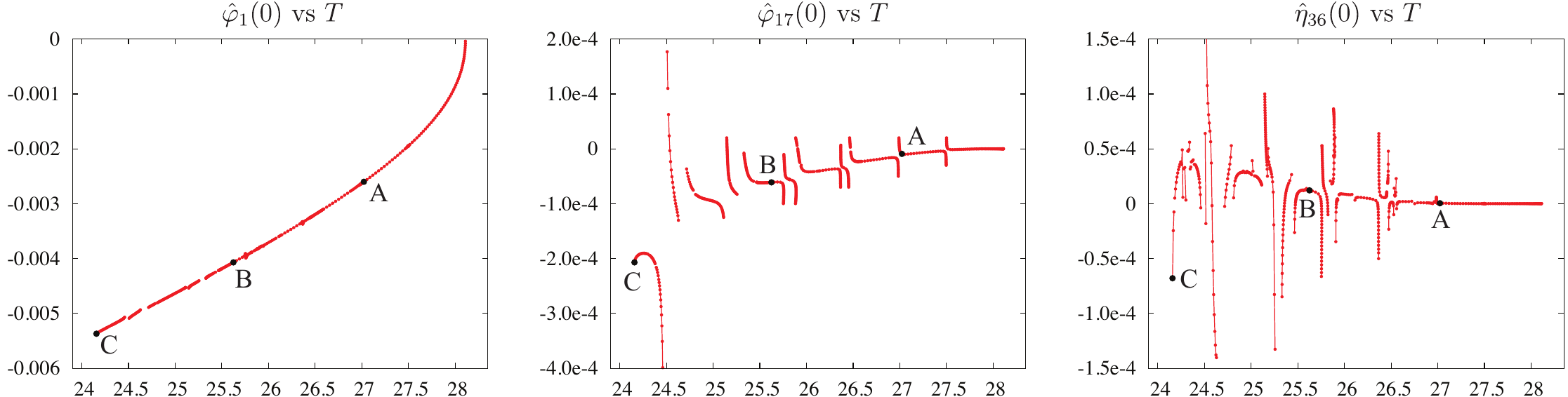}
\end{center}
\caption{\label{fig:bif:005} Bifurcation diagrams showing the
  dependence of $\hat\varphi_1(0)$, $\hat\varphi_{17}(0)$, and
  $\hat\eta_{36}(0)$ on $T$ for a family of standing water waves
  in shallow ($h=0.05$) water. Many more disconnections are
  visible at this depth than were observed in the $h=1.0$ and
  $h=\infty$ cases above.
}
\end{figure}

In Figure~\ref{fig:bif:005}, we track the $k=1$ family of standing
waves out of the linear regime for water of depth $h=0.05$ and spatial
period $2\pi$.  The period of the solutions in the linear regime is
$T=2\pi/\sqrt{\tanh 0.05}=28.1110$, compared to $T=7.19976$ when $h=1$
and $T=2\pi=6.28319$ when $h=\infty$.  Thus, the waves travel much
slower in shallow water.  We also see that $T$ decreases with
amplitude as the waves leave the linear regime, consistent with
Tadjbakhsh and Keller's result, Equation (\ref{eq:omega:correction})
above, that the sign of the quadratic correction term in angular
frequency is positive for $h<1.0581$.  Many more disconnections have
appeared in the bifurcation diagrams at this depth than were observed
in the cases $h\approx 1.0$ and $h=\infty$ considered above.  It was
not possible to track all the side branches that have emerged to see
if they reconnect with each other. However, each time we detected that
the minimization algorithm had jumped from one branch to another, we
did backtrack to fill in enough points to observe which modes were
excited by the resonance.
In general, higher-frequency Fourier modes of the initial condition
possess more disconnections, even though all the bifurcation curves
describe the same family of solutions. For example, in
Fig.~\ref{fig:bif:005}, we see that $\hat\varphi_{17}(0)$ has much
stronger disconnections than $\hat\varphi_1(0)$, and those of
$\hat\eta_{36}(0)$ are stronger still.  This suggests that high-frequency
resonances have little effect on the dynamics of lower-frequency modes.
Nevertheless, there is \emph{some} effect, since even
$\hat\varphi_1(0)$ exhibits visible disconnections, and in fact has
some gaps in $T$ where solutions could not be found. We interpret
these gaps as numerical manifestations of the Cantor-like structures
that arise in analytical studies of standing water waves due to small
divisors \cite{plotnikov01, iooss05}. This will be discussed further
in the conclusion section.

\begin{figure}
\begin{center}
\includegraphics[width=\linewidth]{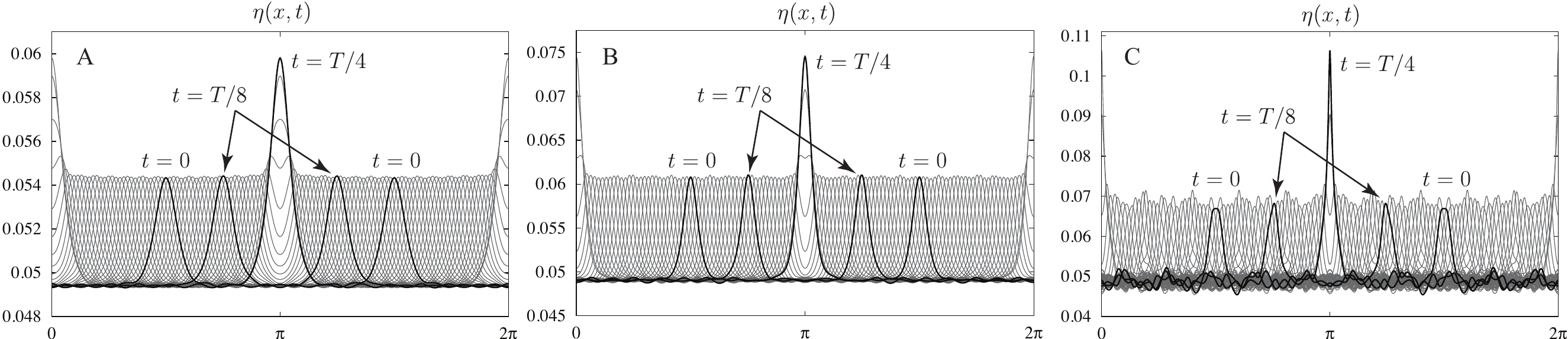}
\end{center}
\caption{\label{fig:evol:005} Standing waves in shallow water take the
  form of counter-propagating solitary waves that interact
  elastically. The low-amplitude radiation normally associated with
  inelastic collisions is already present before the interaction, and
  does not increase as a result of the interaction. In solutions A and
  B, this radiation consists of small-amplitude, high-frequency
  standing waves over which the solitary waves travel. In solution C,
  this radiation is a chaotic mix of standing and counter-propagating
  traveling waves of different wave numbers.  }
\end{figure}

In Figure~\ref{fig:evol:005}, we show time-elapsed snapshots of the
standing wave solutions labeled A--C in Figure~\ref{fig:bif:005}.
These standing waves no longer lead to large scale sloshing modes in
which the fluid rushes from center to sides and back in bulk.
Instead, a pair of counter-propagating solitary waves travel back and
forth across the domain, alternately colliding at $x=\pi$ and $x=0$ at
times $t=T/4+(T/2)\mathbb{Z}$.  In the unit depth case above, we
observed in Figure~\ref{fig:bif100} that disconnections in the
bifurcation curve correspond to secondary standing waves appearing
with one of two phases at the surface of a primary carrier wave.  The
same is true of these solitary wave interactions.  While it is
difficult to observe in a static image, movies of solutions A and B in
Figure~\ref{fig:evol:005} reveal that the primary solitary waves
travel over smaller standing waves with higher wave number and angular
frequency.  As a result, a fluid particle at $x=\pi$ will oscillate up
and down with the secondary standing wave until the solitary waves
collide, pushing the particle upward a great distance. By contrast, in
the infinte-depth case, we saw in Figure~\ref{fig:tip} that
$\eta(\pi,t)$ increases monotonically from trough to crest in spite of
the secondary waves.  Each time a disconnection in
Figure~\ref{fig:bif:005} is crossed, the background standing wave (or
some of its component waves) change phase by $180^\circ$.  Solution B
is positioned near the center of a bifurcation branch, far from major
disconnections in the bifurcation curves. As a result, the water
surface over which the solitary waves travel remains particularly calm
for solution B.  By contrast the background waves of solution C are
quite large in amplitude, with many active wave numbers.  A Floquet
stability analysis, presented elsewhere \cite{water:stable}, shows
that solutions A and B are linearly stable to harmonic perturbations
while solution C is unstable.

\subsection{Gravity-capillary standing waves}
\label{sec:surf}

In this section, we consider the effect of surface tension on the
dynamics of standing water waves.  We restrict attention to waves of
the type considered by Concus \cite{concus:62} and Vanden-Broeck
\cite{vandenBroeck:84}, leaving collisions of gravity-capillary waves
\cite{milewski:11} for future work \cite{water:stable}.
The only change in the linearized equations (\ref{eq:lin}) when
surface tension is included is that
$\dot\varphi_t=P\big[-g\dot\eta+(\sigma/\rho)\dot\eta_{xx}\big]$.  The
standing wave solutions of the linearized problem continue to have the
form (\ref{eq:lin:soln}), but with
\begin{equation}
  \omega^2 = \left(g + \frac{\sigma}{\rho}k^2\right)k\tanh kh, \qquad
  A/B = \sqrt{k\tanh kh\Big/\big[g + (\sigma/\rho)k^2\big]}.
\end{equation}
We choose length and time-scales so that $g=1$ and $\sigma/\rho=1$.
For simplicity, we consider only the $k=1$ bifurcation in the infinite
depth case, and continue to assume all functions are $2\pi$-periodic
in space. In this configuration, the period of the linearized standing
waves is $T=2\pi/\sqrt{2}\approx 4.443$.  For real water (assuming
$\sigma=72 \text{ dyne}/\text{cm}$), 4.443 units of dimensionless
time corresponds to $0.0739$ seconds, and $2\pi$ spatial units
corresponds to $1.7$ cm.

\begin{figure}
\begin{center}
\includegraphics[width=\linewidth]{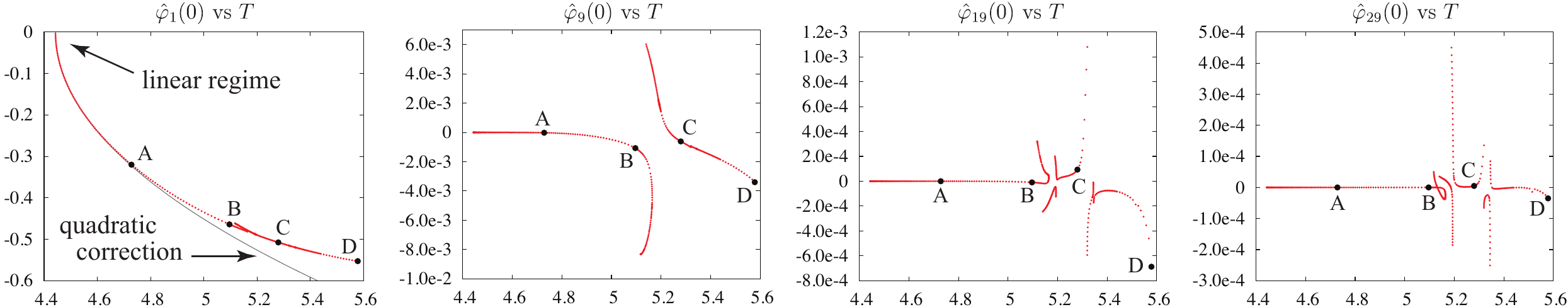}
\end{center}
\caption{\label{fig:bif:surf} Bifurcation diagrams showing the
  dependence of various Fourier modes of the initial conditions on the
  period for standing water waves with surface tension.  The bifurcation
  curve splits into several disjoint branches between solutions B and C
  as resonant waves appear on the fluid surface. The curve labeled
  `quadratic correction' is given in Equation (\ref{eq:concus}).
}
\end{figure}

\begin{figure}
\begin{center}
\includegraphics[width=\linewidth]{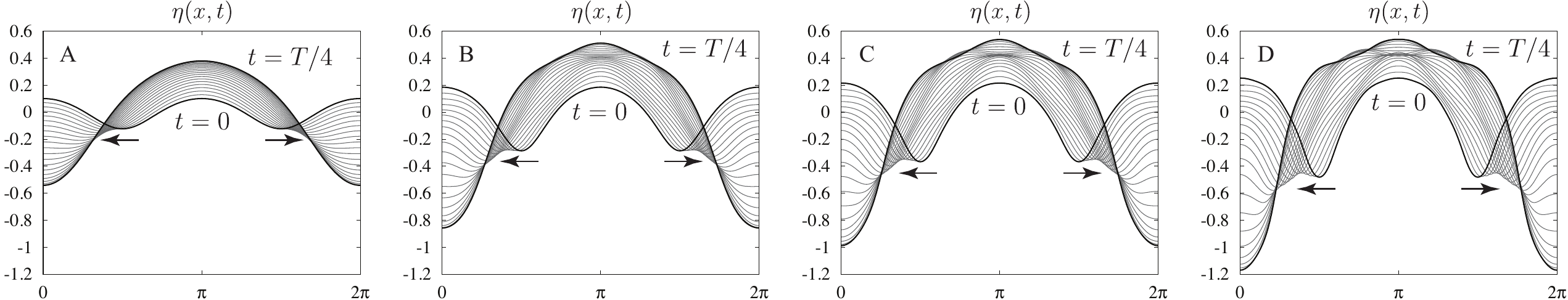}
\end{center}
\caption{\label{fig:evol:surf} Time-elapsed snapshots of four standing
  waves over a quarter-period.  At $t=0$, a pair of
  counter-propagating depression waves move away from each other as
  the fluid flows to the center.  Solutions B, C and D exhibit
  higher-frequency standing waves oscillating on the surface of the
  low-frequency carrier wave. All of the solutions reach a rest state
  where $\varphi\equiv0$ at $t=T/4$. }
\end{figure}

\begin{figure}
\begin{center}
\includegraphics[width=\linewidth]{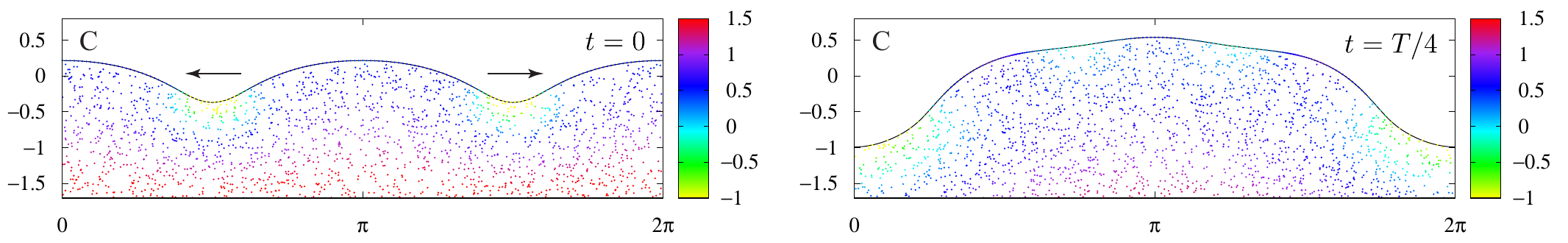}
\end{center}
\caption{\label{fig:evol:surf:P} A more detailed view of solution C
  from Figure~\ref{fig:evol:surf} at times $t=0$ and $t=T/4$ showing
  regions of negative pressure beneath depression waves. These images
  are taken from movies in which passively advected particles have
  been added to the fluid for visualization, color coded by pressure
  using (\ref{eq:pressure}).  A secondary standing wave leads to
  visible variations in curvature and pressure at time $T/4$.  }
\end{figure}

The results are summarized in Figure~\ref{fig:bif:surf}. As the
bifurcation parameter, $c_1=\hat\varphi_1(0)$, increases in magnitude
beyond the realm of linear theory, the period increases, just as in
the zero surface tension case for $h=\infty$. Quantitatively, our
results agree with Concus' prediction \cite{concus:62} that
\begin{equation}\label{eq:concus}
  T = \sqrt{2}\,\pi\left(1 + \frac{197}{320}c_1^2\right), \qquad
  c_1 = \hat\varphi_1(0)
\end{equation}
in the infinite depth case when the surface tension parameter
$\delta:=\sigma k^2/(\sigma k^2+\rho g)$ is equal to 1/2.  Equation
(\ref{eq:concus}) is plotted in the left panel of
Figure~\ref{fig:bif:surf} for comparison.
Shortly after solution B ($c_1=-0.464$, $T=5.10$), a complicated
sequence of imperfect bifurcations occurs in which several disjoint
families of solutions pass near each other. Comparison of solutions B,
C and D in Figure~\ref{fig:evol:surf} suggests that these
disconnections are due to the excitation of different patterns of
smaller-scale capillary waves oscillating on the free surface.  An
interesting difference between these standing waves and their zero
surface-tension counterparts (e.g.~in Figure~\ref{fig:bif10}) is that
the ``solitary'' waves that appear in the transition periods between
rest states of maximum amplitude are inverted.  Thus, we can think of
these solutions as counter-propagating depression waves \cite{crapper,
  vandenBroeck:crapper} that are tuned to be time-periodic, just as
the zero surface-tension case leads to counter-propagating Stokes
waves.  The depression waves travel outward as fluid flows to the
center, whereas the Stokes waves travel inward, carrying the fluid
with them.  Figure~\ref{fig:evol:surf:P} shows snapshots of particle
trajectories for solution C, color coded by pressure. The methodology
for computing this pressure is given at the end of
Appendix~\ref{sec:BI}.  Negative pressure (relative to the ambient air
pressure $p_0$ in (\ref{eq:c:def}), which is set to zero for
convenience) arises beneath the depression waves as they pass, which
leads to larger pressure gradients, faster wave speeds, and shorter
periods than were seen in previous sections.

\subsection{Performance comparison}
\label{sec:perform}

We conclude our results with a comparison of running times for
the various algorithms and machines used to generate the data reported
above.  Our machines consist of a laptop, a desktop, a server, a GPU
device, and a supercluster.  The laptop is a Macbook Pro, 2.53 GHz
Intel Core i5 machine. The desktop is a Mac Pro with two quad-core 2.8
GHz Intel Nehalem processors.  The rackmount server has two six-core
3.33 GHz Intel Westmere processors and an NVidia M2050 GPU, and is
running Ubuntu Linux. The cluster is the Lawrencium cluster (LR1) at
Lawrence Berkeley National Laboratory.  Each node of the cluster
contains two quad-core 2.66 GHz Intel Harpertown processors.  Intel's
math kernel library and scalapack library were used for the linear
algebra on Lawrencium.

\begin{table}[b]
\begin{equation*}
  \begin{array}{c|c|c|c|c|c|c|c|c|c}
    \text{index range} & n & M & N & N_\text{quad} & \text{bif par} & \text{start} & \text{end} & T_\text{start} & T_\text{end} \\ \hline
    \text{101--150} & 20 &  128 &  60 &  24 & \hat\varphi_1(0) & -0.004 & -0.2   & 1.5708 & 1.6034 \\
    \text{150--174} & 32 &  192 &  96 &  36 & \hat\varphi_1(0) & -0.2   & -0.26  & 1.6034 & 1.6265 \\
    \text{174--184} & 50 &  256 &  96 &  48 & \hat\varphi_1(0) & -0.26  & -0.275 & 1.6265 & 1.6332 \\
    \text{184--194} & 54 &  384 & 120 &  60 & \hat\varphi_5(0) &  0.001071 & 0.001856   & 1.6332 & 1.6359 \\
    \text{194--200} & 64 &  512 & 144 &  72 & \hat\varphi_5(0) &  0.001856 & 0.002117   & 1.6359 & 1.6358 \\
    \text{200--210} & 75 &  768 & 180 &  96 & \hat\varphi_5(0) &  0.002117 & 0.002515   & 1.6358 & 1.6348 \\
    \text{210--220} & 96 & 1024 & 240 & 120 & \hat\varphi_5(0) &  0.002515 & 0.002981   & 1.6348 & 1.6326
  \end{array}
\end{equation*}
\caption{\label{tab:param:small}
  Parameters used in the performance comparison for small problems.
  Here ``start'' and ``end'' give the values of the bifurcation
  parameter (bif par) at the endpoints of the corresponding segment
  of the bifurcation curve.  The solutions at these endpoints (with
  index 150, 174, 184, etc.) are computed twice, once on the coarse mesh
  and once on the fine mesh.  $T_\text{start}$ and $T_\text{end}$ are
  the periods at the endpoints.  $N_\text{quad}$ is the number of
  timesteps (of the SDC scheme) used in the quadruple precision
  calculations.  $M_\text{quad}$ and $n_\text{quad}$ were set equal to
  $2M$ and $2n$, respectively.
}
\end{table}

\begin{figure}
\begin{center}
\includegraphics[width=\linewidth]{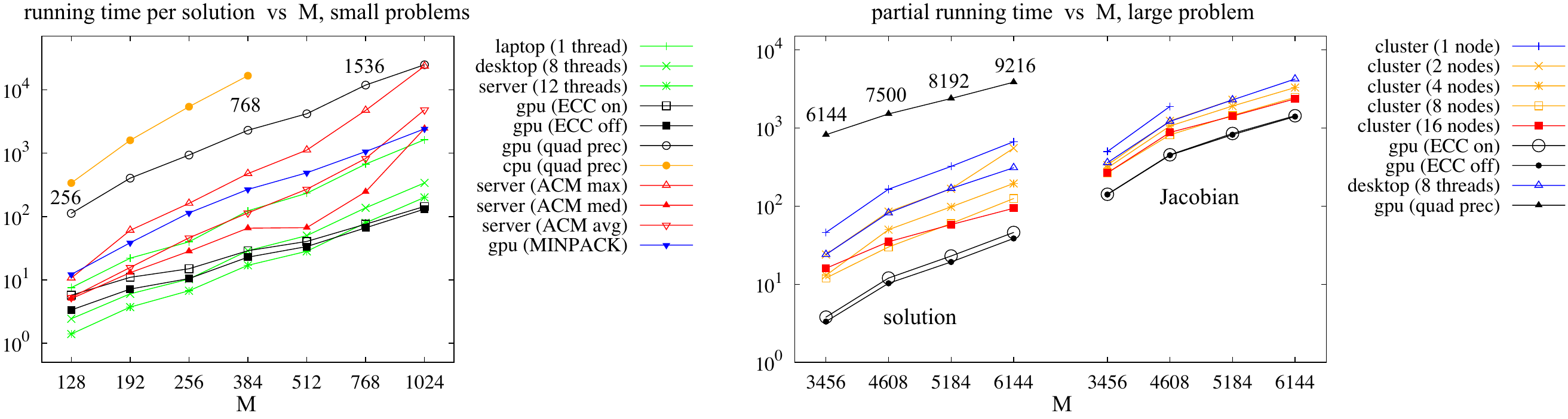}
\end{center}
\caption{\label{fig:times} Performance of the algorithms on various
  architectures. (left) Each data point is the average running time
  (in seconds per solution) for solutions in each index range listed
  in Table~\ref{tab:param:small}.  For the Adjoint Continuation Method
  (ACM), which takes much longer for the first solution than
  subsequent solutions due to re-use of the Hessian information, we
  also report the longest running time and the median running
  time. (right) Time taken in each segment of the mesh-refinement
  strategy to evolve solution O in Figure~\ref{fig:breakdown} through
  $1/60$th of a quarter-period.  The parameters for each segment are
  given in Table~\ref{tab:param:large}.  The times listed for the
  Jacobian are the cost of evolving all 1200 columns through time
  $T/240$.  }
\end{figure}

Our first test consists of computing the first 120 deep-water standing
wave solutions reported in Figure~\ref{fig:bif:00} (up through
solution B).  The running times increase with amplitude due to an
increase in the number of gridpoints ($M$), timesteps ($N$), and
unknown Fourier modes of the initial conditions ($n$).  The parameters
used in this test are given in Table~\ref{tab:param:small}, with running
times reported in the left panel of Figure~\ref{fig:times}.  For
each index range, we computed the average time required to reduce $f$
below $10^{-25}$ (or $10^{-50}$ in quadruple precision), using linear
extrapolation from the previous two solutions as a starting guess. In
the Adjoint Continuation Method, the first solution in each range
(with index 101, 150, 174, etc.) takes much longer than subsequent
minimizations. This is because we re-build the inverse Hessian from
scratch when the problem size changes, but not from one solution to
the next in a given index range. This is illustrated in the figure by
plotting the maximum and median number of seconds required to find a
solution in a given index range, along with the average.

For these smaller problems, the DOPRI5 and DOPRI8 schemes are of
comparable efficiency for double-precision accuracy.  We used the
former for this particular test.  In quadruple precision, we switched
to the SDC15 scheme, which is more efficient than DOPRI5 and DOPRI8 in
reducing $f$ below $10^{-50}$.  We also doubled $M$ and $n$ in the
quadruple-precision runs.  The MINPACK benchmark results were
optimized as much as possible (using the GPU with Error Correcting
Code (ECC) turned off) to give as fair a comparison as possible. For
the benchmark, the Jacobian was computed via forward differences, as
in (\ref{eq:J:fd}).  The ACM method works well on small problems, but
starts to slow down relative to the benchmark around $M=1024$.  The
trust region shooting method is much faster than the ACM (and the
benchmark) due to the fact that all the columns of the Jacobian employ
the same Dirichlet to Neumann operator at each timestep. Thus, we save
a factor of $n$ in setup costs by computing $n$ columns of the
Jacobian simultaneously. Moreover, most of the work can be organized
to run at level 3 BLAS speed.  Note that the GPU is slower than the
multi-core CPU in double-precision for small problems, but eventually
wins out as the opportunity for parallelism increases.  In
quadruple-precision, the GPU is substantially faster than the CPU for
all problem sizes tested as there is more arithmetic to be done
relative to communication costs.

\begin{table}[t]
\begin{equation*}
  \begin{array}{r||c||c|c|c|c||c|c|c|c||c||c|c|c|c||c|c|c|c||c}
    & \nu & \theta_1 & \theta_2 & \theta_3 & \theta_4 & \rho_1 & \rho_2 & \rho_3 & \rho_4 & n & M_1 & M_2 & M_3 & M_4 &  d_1 & d_2 & d_3 & d_4 & \text{scheme} \\ \hline
    \text{double} & \js 4 & \js 0.2 & \js 0.2 & \js 0.2 & \js 0.4 & \js 1.0 & \js 0.4 & \js 0.12 & \js 0.09 & \js 1200 & \js 3456 & \js 4608 & \js 5184 & \js 6144 & \js 10 & \js 20 & \js 30 & \js 40 & \text{\footnotesize DOPRI8} \\
    \text{quad} & \js 4 & \js 0.1 & \js 0.3 & \js 0.4 & \js 0.2 & \js 1.0 & \js 0.25 & \js 0.08 & \js 0.05 & \js 1500 & \js 6144 & \js 7500 & \js 8192 & \js 9216 & \js  10 & \js 12 & \js 16 & \js 20 & \text{\footnotesize SDC15}
  \end{array}
\end{equation*}
\caption{ \label{tab:param:large}
  Parameters used in the performance comparison for a large problem
  (solution O of Figure~\ref{fig:breakdown}).  Here $d_l$ is the number
  of timesteps to advance the solution through one sixtieth of a quarter
  period ($T/240$).  The total number of timesteps on segment $l$ is
  $N_l = 60\theta_l d_l$ in this case, while the number of function
  evaluations is $12N_l$ for DOPRI8 and $99N_l$ for SDC15. 
}

\end{table}

Our second test consists of timing each phase of the computation of
solution O in Figure~\ref{fig:breakdown}.  The parameters used for the
performance comparison are given in Table~\ref{tab:param:large}.  For
the double-precision calculation, we later refined the mesh to the
values listed in Table~\ref{tab:param:OA} in
Section~\ref{sec:breakdown}; however, this was done on one machine
only. (The value of $f$ here is $3.9\times 10^{-23}$ versus
$1.3\times10^{-26}$ in Section~\ref{sec:breakdown}.)  The quantities
$d_l$ in Table~\ref{tab:param:large} are the number of timesteps
between mile-markers where the energy and plots of the solution were
recorded.  In this case, we recorded 60 slices of the solution between
$t=0$ and $t=T/4$.  The running times in the right panel of
Figure~\ref{fig:times} report the time to advance from one mile-marker
to the next.  It was not possible to solve this problem via the ACM
method or MINPACK, so this test compares running times of the trust
region method on several machines.  In quadruple precision, we evolved
the solution but did not compute the Jacobian.  Two of the jobs on the
cluster (1 node and 2 nodes) were terminated early due to insufficient
available wall-clock time.  When using the GPU, there is little
improvement in performance in also running openMP on the CPU.  For
example, switching from 12 threads (shown in the figure) to one thread
(not shown) slows the computation of the Jacobian by about 10 percent,
but speeds up the computation of the solution by about 1 percent.
When evolving the solution on a large problem, the GPU is fully
utilized; however, when evolving the Jacobian, the GPU is idle about
60 percent of the time.  Thus, we can run 2-3 jobs simultaneously to
improve the effective performance of the GPU by another factor of 2
over what is plotted in the figures.  This is also true of the
Lawrencium cluster --- while using more nodes to solve a single problem
stops paying off around 8 nodes, we can run multiple jobs
independently. Most of the large-amplitude solutions in
Figure~\ref{fig:bif:00} were computed in this way on the Lawrencium
cluster, before we acquired the GPU device.

\section{Conclusion}
\label{sec:conclude}

We have shown how to compute time-periodic solutions of the
free-surface Euler equations with improved resolution, accuracy and
robustness by formulating the shooting method as an overdetermined
nonlinear least squares problem and exploiting parallelism in the
Jacobian calculation.  This made it possible to resolve a
long-standing open question, posed by Penney and Price in 1952, on
whether the most extreme standing wave develops wave crests with sharp
90 degree corners each time the fluid comes to rest.  Previous
numerical studies reached different conclusions about the form of the
limiting wave, but none were able to resolve the fine-scale
oscillations that develop due to resonant effects.  While we cannot
say for certain that no standing wave exists that forms sharp corners
at periodic time-intervals, we can say that such a wave does not lie
at the end of a family of increasingly sharp standing waves
parametrized by crest acceleration, $A_c$.  Indeed, crest acceleration
is not a monotonic function, and the bifurcation curve becomes
fragmented as $A_c\rightarrow1$, with different branches corresponding
to different fine-scale oscillation patterns that emerge at the
surface of the wave.  Following any of these branches in either
direction leads to increasingly oscillatory solutions with curvature
that appears to blow up throughout the interval $[0,2\pi]$, not just
at the crest tip.

Small-amplitude standing waves have been proved to exist in finite
depth by Plotnikov and Toland \cite{plotnikov01}, and in infinite
depth by Plotnikov, Toland and Iooss \cite{iooss05}.  However, the
proofs rely on a Nash-Moser iteration that only guarantees existence
for values of the amplitude in a totally disconnected Cantor set
\cite{craig:wayne,bourgain99}.
In shallow water, with $h=0.05$, we do see evidence that solutions do
not come in smooth families. For example, in Figure~\ref{fig:bif:005},
the number of visible disconnections in the bifurcation diagrams
increases dramatically from $\hat\varphi_1(0)$ to
$\hat\varphi_{17}(0)$ to $\hat\eta_{36}(0)$.
There are also a few gaps along the $T$-axis where the numerical
method failed to find a solution, i.e.~the minimum value of $f$ did
not decrease below the target of $10^{-26}$ regardless of how many
additional Fourier modes were included in the simulation.  It is easy
to imagine that removing all the gaps that arise in this fashion as
the mesh is refined and the numerical precision is increased could
lead to a Cantor-like set of allowed periods.

Our numerical method measures success by how small the objective
function $f$ and residual $r$ become.  It will succeed if it can find
initial conditions that are close enough to those of an exactly
time-periodic solution, or at least of a solution that is time-periodic
up to roundoff error tolerances.
For the residual to be small, the bifurcation parameter must
nearly belong to the Cantor set of allowed values, 
but membership need not be exact.  If the Cantor set is fat enough
(i.e.~has nearly full measure), then most values of the bifurcation
parameter will be close to some element of the set --- roundoff error
fills in the smallest gaps.  While it is possible that our numerical
method would report a false positive, this seems unlikely.
The residuals of our solutions are not under-predicted by the
minimization algorithm due to formulation of the problem as an
overdetermined system.  Indeed, we saw in Figure~\ref{fig:evol:PhiOA}
that $f$ decreases from $1.3\times10^{-26}$ to $8.6\times10^{-27}$ for
solution O when the initial conditions are evolved on a finer mesh in
quadruple precision, and decreases from $1.9\times10^{-28}$ to
$2.1\times10^{-60}$ for solution A when the minimization is repeated
in quadruple precision. This latter test is particularly convincing
that the method is converging to an exactly time-periodic solution.

\begin{figure}[b]
\begin{center}
\includegraphics[width=\linewidth]{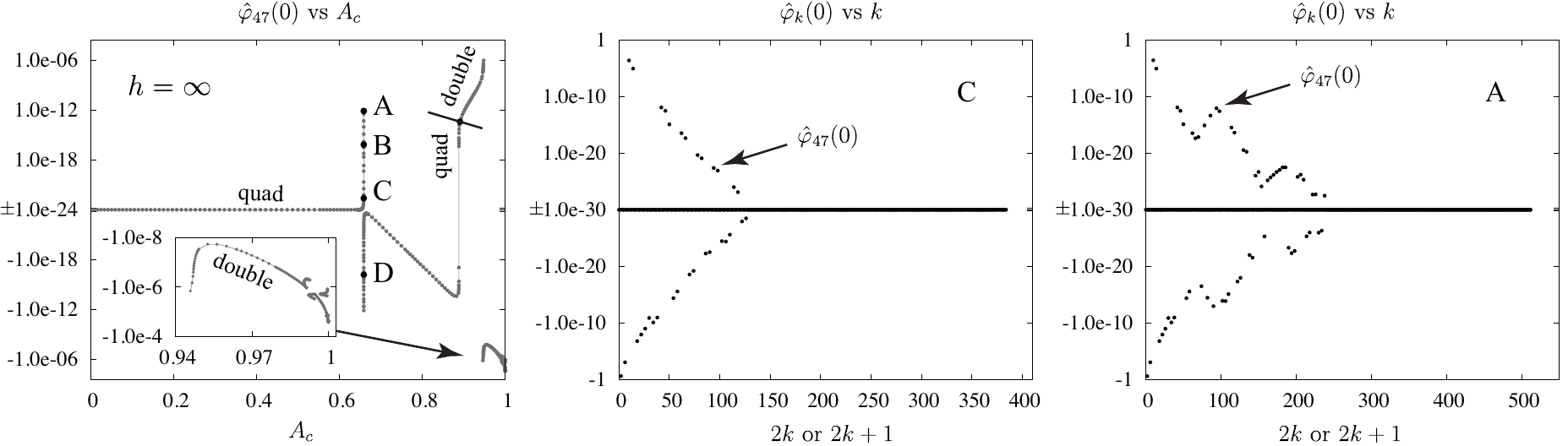}
\end{center}
\caption{\label{fig:quad:00} Study of resonance in deep water standing
  waves. (left) Switching from double- to quadruple-precision
  arithmetic reveals only one additional disconnection in the
  bifurcation curves. The inset graph shows how the disconnections of
  Figure~\ref{fig:bif:00} look when $c_{47}$ is plotted rather than
  $c_1$, $c_5$ and $c_{60}$.  Solutions B and D are the points where
  $|c_{47}/c_1|\approx 3\times 10^{-16}$, just barely above the
  roundoff threshold.  The gap in crest acceleration between these
  solutions is $A_c(D)-A_c(B)=1.4\times 10^{-9}$; thus, it is
  extremely unlikely in a parameter study that one would land in this
  gap. Outside of this gap, resonant effects from this disconnection
  are smaller than the roundoff threshold.  (center and right)
  Resonance causes bursts of growth in the Fourier spectrum, but the
  modes continue to decay exponentially in the long run. }
\end{figure}

If standing waves on water of infinite depth do not come in smooth
families, as suggested by the analysis of \cite{iooss05}, they are
remarkably well approximated by them.  Prior to our work, no numerical
evidence of disconnections in the bifurcation curves had been
observed.  Wilkening \cite{breakdown} found several disconnections for
values of crest acceleration $A_c>0.99$, but none at smaller values.
As shown in Figure~\ref{fig:quad:00}, there is one additional
disconnection around $A_c=0.947$ that can be observed in
double-precision that was missed in \cite{breakdown}.  However, the
points at which resonance is supposed to cause difficulty are expected
to be dense over the whole range $0<A_c<1$.  We re-computed the
solutions up to $A_c=0.8907$ in quadruple precision, expecting several
new disconnections to emerge in high-frequency Fourier modes.
Surprisingly, we could only find one, at $A_c=0.658621$. Using a
bisection algorithm to zoom in on the disconnection in the 47th
Fourier mode from both sides (using $c_5$ as the bifurcation
parameter), we were able to extend the side branches from
$c_{47}\approx \pm 10^{-27}$ to $c_{47}\approx \pm 10^{-12}$.  These
side branches become observable in double-precision at points B and D
in Figure~\ref{fig:quad:00}.  However, the gap in crest acceleration
between solutions B and D is only $1.4\times 10^{-9}$ units wide.
Thus, it is extremely unlikely that this resonance could be detected
in double-precision without knowing where to look.  Presumably the
same issue prevents us from seeing additional disconnections in
quadruple-precision.  This suggests that the Cantor-like set of
allowed values of the amplitude parameter is very fat, with gaps
decaying to zero rapidly with the wave number of the resonant mode.

\begin{figure}
\begin{center}
\includegraphics[width=\linewidth]{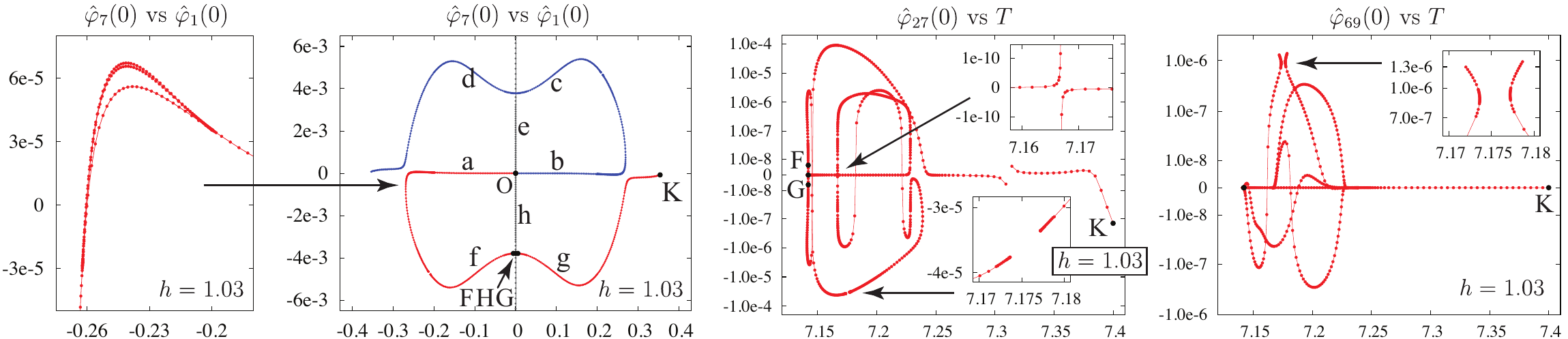}
\end{center}
\caption{\label{fig:bif103w} A closer look at the bifurcation structure
  in Figure~\ref{fig:spider} in the $h=1.03$ case reveals a number of
  additional side-branches that trace back and forth over nearly the
  same curves when low-frequency modes are plotted (left), but become
  well-separated when high frequency modes are plotted (right center,
  right). A small gap
  near $T=7.175$ has formed on one of the wings in the plots of
  $\hat\varphi_{27}(0)$ and $\hat\varphi_{69}(0)$ vs $T$.
}
\end{figure}

In finite depth, with $h\approx 1$, a connection can be seen between
resonance and non-uniqueness of solutions.  The main difference from
the $h=0.05$ and $h=\infty$ cases is that for $h\approx1$, the
disconnections lead to side-branches that can be tracked a great
distance via numerical continuation, and are often found to be
globally connected to one another.
%, just as in the vortex sheet with surface tension \cite{vtxs1}.
Traversing these side branches causes high-frequency modes to sweep
out small-amplitude loop-shaped structures.  These loops are ``long
and thin'' in the sense that low-frequency modes trace back over the
previously swept out bifurcation curves while traversing the loop,
with little deviation in the lateral direction.  For example, in
Figure~\ref{fig:bif103w}, the 27th Fourier mode
executes a number of excursions in which it grows to around $10^{-6}$,
causing the period and first Fourier modes to sweep back and forth
over much larger ranges, $7.167<T<7.229$ and
$-0.200>\hat\varphi_1(0)>-0.262$.  These loops are plotted in
Figures~\ref{fig:spider} and~\ref{fig:bif104} as well, but the
curves are indistinguishable from one another at this resolution since
the lateral deviations are so small.  Looking at the third panel of
Figure~\ref{fig:bif103w}, one might ask, ``how many solutions are
there with period $T=7.2$.''  If we had not noticed any of the
disconnections (note the exponential scaling of the axis), we would
have answered 1.  If we had only tracked the outer wings, we would
have answered 3. Having tracked all the branches shown, the answer
appears to be 5.  But of course there are probably infinitely many
disconnections in higher-frequency Fourier modes that we did not
resolve or track, and some of these may lead to additional
solutions with $T=1.2$.  Physically, all these crossings of $T=1.2$
correspond to a hierarchy of ``standing waves on standing waves,''
with different mode amplitudes and phases working together to create a
globally time-periodic solution with this period.  The fact that the
low-frequency bifurcation curves sweep back and forth over nearly the
same graph reinforces the physically reasonable idea that
high-frequency, low amplitude waves oscillating on the surface of
low-frequency, large amplitude waves will not significantly change the
large-scale behavior.

In summary, time-periodic water waves occur in abundance in numerical
simulations, and appear to be highly non-unique, partly due to the
Wilton's ripple phenomenon of mixed-mode solutions co-existing with
pure-mode solutions near a degenerate bifurcation, and also due to a
tendency of the bifurcation curves to fold back on themselves each
time a resonant mode is excited.  Proofs of existence based on
Nash-Moser iteration must somehow select among these multiple
solutions, and it would be interesting to know whether the Cantor-like
structure in the analysis is caused by a true lack of existence for
parameter values outside of this set, or is partly caused by
non-uniqueness.  Finally, we note that most of the disconnections in
the numerically computed bifurcation curves disappear in the infinite
depth limit, and remarkably small residuals can be achieved with
smooth families of approximate solutions. This calls for further
investigation of the extent to which the obstacles to proving smooth
dependence of solutions on amplitude can be overcome or quantified.

\newpage

\vspace*{.25in}
\noindent\textbf{Acknowledgments}
\vspace*{.125in}

\noindent This research was supported in part by the Director, Office of
Science, Computational and Technology Research, U.S.  Department of
Energy under Contract No.~DE-AC02-05CH11231, and by the National
Science Foundation through grant DMS-0955078. Some of the computations
were performed on the Lawrencium supercluster at LBNL.

\appendix

\section{Boundary integral formulation}
\label{sec:BI}

While many numerical methods exist to evolve irrotational flow
problems \cite{lh76, baker:82, krasny:86, mercer:92, dias:bridges,
  HLS01, vtxs1, dyachenko:1996, milewski:11, baker10}, we have found that a
direct boundary integral implementation of (\ref{eq:ww}) is the
simplest and most effective approach for problems where $\eta$ remains
single valued, i.e.~the interface does not overturn.  Suppressing $t$
in the notation, we represent the complex velocity potential
$\Phi(z)=\phi(z)+i\psi(z)$ as a Cauchy integral \cite{muskh}
\begin{equation}\label{eq:cauchy}
  \Phi(z) = \frac{1}{2\pi i}PV\!\!\int_{-\infty}^\infty
  \frac{-\zeta'(\alpha)}{\zeta(\alpha)-z}\mu(\alpha)\,d\alpha, \qquad
  \zeta(\alpha) = \xi(\alpha) + i\eta(\xi(\alpha)),
\end{equation}
where $z$ is a field point in the fluid, $\mu(\alpha)$ is the
(real-valued) dipole density, $\zeta(\alpha)$ parametrizes the free
surface, $PV$ indicates a principal value integral, $\eta(x)$
retains its meaning from equation (\ref{eq:ww}), and the change of
variables $x=\xi(\alpha)$ will be used to smoothly refine the mesh in
regions of high curvature. The minus sign in (\ref{eq:cauchy})
accounts for the fact that Cauchy integrals are usually parametrized
counter-clockwise, but we have parametrized the curve so the fluid
region lies to the right of $\zeta(\alpha)$. 
When the fluid depth is finite, we impose the tangential flow
condition using an identical double-layer potential on the mirror
image surface, $\bar\zeta(\alpha)$. This assumes we have set $h=0$ in
(\ref{eq:dno}), absorbing the mean fluid depth into $\eta$ itself.  We
also use $\frac{1}{2}\cot\frac{z}{2} = PV\sum_k\frac{1}{z+2\pi k}$ to
sum (\ref{eq:cauchy}) over periodic images.  The result is
\begin{equation}\label{eq:Phi}
  \Phi(z) = \frac{1}{2\pi i}\int_0^{2\pi}
  \left[ \frac{\zeta'(\alpha)}{2}\cot\left(\frac{z-\zeta(\alpha)}{2}
        \right) -
        \frac{\bar\zeta'(\alpha)}{2}\cot\left(\frac{z-\bar\zeta(\alpha)}{2}
        \right)\right]\mu(\alpha)\,d\alpha.
\end{equation}
Note that $\Phi$ is real-valued on the $x$-axis, indicating that the
stream function $\psi$ is zero (and therefore constant) along the bottom
boundary.

As $z$ approaches $\zeta(\alpha)$ from above ($+$) or below ($-$), the
Plemelj formula \cite{muskh} gives
\begin{equation}\label{eq:plemelj:1}
  \Phi\big(\zeta(\alpha)^\pm\big) = \mp\frac{1}{2}\mu(\alpha) +
  \frac{PV}{2\pi i}\int_0^{2\pi} 
  \left[
    \frac{\zeta'(\beta)}{2}\cot\frac{\zeta(\alpha)-
      \zeta(\beta)}{2} -
  \frac{\bar\zeta'(\beta)}{2}\cot\frac{\zeta(\alpha)-
      \bar\zeta(\beta)}{2}
    \right]
  \mu(\beta)\,d\beta.
\end{equation}
We regularize the principal value integral by subtracting and adding
$\frac{1}{2}\cot\left(\frac{\alpha-\beta}{2}\right)$ from the first
term in brackets \cite{vande:vooren:80,pullin:82,baker:nachbin:98}.
The result is
\begin{equation}
  \label{eq:Phi:formula}
  \Phi\big(\zeta(\alpha)^\pm\big) = \mp\frac{1}{2}\mu(\alpha)
  - \frac{i}{2}H\mu(\alpha) + \frac{1}{2\pi i}\int_0^{2\pi}
  [\wtil K_1(\alpha,\beta) + \wtil K_2(\alpha,\beta)]
  \mu(\beta)\,d\beta,
\end{equation}
where $Hf(\alpha)=\frac{1}{\pi}PV\!\int_{-\infty}^\infty
\frac{f(\beta)}{\alpha-\beta}\,d\alpha = \frac{1}{\pi}PV\!
\int_0^{2\pi}\frac{f(\beta)}{2}\cot\left(
  \frac{\alpha-\beta}{2}\right)d\beta$ is the Hilbert transform and
\begin{equation}
  \label{eq:K1K2:til}
  \wtil K_1(\alpha,\beta) = \frac{\zeta'(\beta)}{2}
  \cot\frac{\zeta(\alpha) - \zeta(\beta)}{2}
  - \frac{1}{2}\cot\frac{\alpha-\beta}{2}, \qquad
  \wtil K_2(\alpha,\beta) = \frac{\bar\zeta'(\beta)}{2}\cot
  \frac{\zeta(\alpha)-\bar\zeta(\beta)}{2}.
\end{equation}
We note that $\wtil
K_1(\alpha,\beta)$ is continuous at $\beta=\alpha$ if we define $\wtil
K_1(\alpha,\alpha)=-\zeta''(\alpha)/[2\zeta'(\alpha)]$.  Taking the
real part of (\ref{eq:Phi:formula}) at $z=\zeta(\alpha)^-$ yields a
second-kind Fredholm integral equation for $\mu(\alpha)$ in terms of
$\varphi(\xi(\alpha))$,
\begin{equation}\label{eq:fred2}
  \frac{1}{2}\mu(\alpha) + \frac{1}{2\pi}\int_0^{2\pi}
  [K_1(\alpha,\beta) + K_2(\alpha,\beta)]\mu(\beta)\,d\beta =
  \varphi(\xi(\alpha)),
\end{equation}
where $K_j(\alpha,\beta) = \im\{ \wtil K_j(\alpha,\beta)\}$.
Once $\mu(\alpha)$ is known, it follows from (\ref{eq:Phi})
that
\begin{equation}
  \label{eq:dPhi1}
  \Phi'(z) = u(z)-iv(z) = \frac{1}{2\pi i}\int_0^{2\pi}
  \left[\frac{1}{2}\cot\left(\frac{z-\zeta(\alpha)}{2}\right) -
    \frac{1}{2}\cot\left(\frac{z-\bar{\zeta}(\alpha)}{2}\right)\right]
  \gamma(\alpha)\,d\alpha,
\end{equation}
where $\gamma(\alpha)=\mu'(\alpha)$ is the (normalized) vortex sheet
strength.  As $z$ approaches $\zeta(\alpha)$ from above or below, one
may show \cite{jia:thesis} that
\begin{equation}\label{eq:plemelj:2}
  \zeta'(\alpha)\Phi'\big(\zeta(\alpha)^\pm\big) =
  \mp\frac{1}{2}\gamma(\alpha) + \frac{PV}{2\pi i}\int_0^{2\pi}
  \left[
    \frac{\zeta'(\alpha)}{2}\cot\frac{\zeta(\alpha) - \zeta(\beta)}{2}
    - \frac{\zeta'(\alpha)}{2}\cot\frac{\zeta(\alpha) - \bar\zeta(\beta)}{2}
    \right]\gamma(\beta)\,d\beta.
\end{equation}
Note that $\zeta'$ is evaluated at $\beta$ in (\ref{eq:plemelj:1})
and at $\alpha$ in (\ref{eq:plemelj:2}) inside the integral.
We regularize the principal value integral using the same technique
as before to obtain
\begin{equation}
\label{eq:dPhi2}
  \zeta'(\alpha)\Phi'\big(\zeta(\alpha)^\pm\big) =
  \mp\frac{1}{2}\gamma(\alpha) - \frac{i}{2}H\gamma(\alpha) +
  \frac{1}{2\pi i}\int_0^{2\pi} [\wtil G_1(\alpha,\beta)
  + \wtil G_2(\alpha,\beta)]\gamma(\beta)\,d\beta,
\end{equation}
where
\begin{equation}
  \label{eq:G1G2:til}
  \wtil G_1(\alpha,\beta) = \frac{\zeta'(\alpha)}{2}
  \cot\frac{\zeta(\alpha) - \zeta(\beta)}{2}
  - \frac{1}{2}\cot\frac{\alpha-\beta}{2}, \qquad
  \wtil G_2(\alpha,\beta) = \frac{\zeta'(\alpha)}{2}\cot
  \frac{\zeta(\alpha)-\bar\zeta(\beta)}{2}.
\end{equation}
$\wtil G_1(\alpha,\beta)$ is continuous at $\beta=\alpha$ if we define
$\wtil G_1(\alpha,\alpha) = \zeta''(\alpha)/[2\zeta'(\alpha)]$.  We
could read off $u=\phi_x$ and $v=\phi_y$ from (\ref{eq:dPhi2}) for use
in the right hand side of (\ref{eq:ww}).  Instead, as an intermediate
step, we compute the output of the Dirichlet-Neumann operator
defined in (\ref{eq:DNO:def}),
\begin{equation}\label{eq:DNO:formula}
\begin{aligned}
  |\xi'(\alpha)|\mc G\varphi(\xi(\alpha)) &=
  |\zeta'(\alpha)|\der{\phi}{n}(\zeta(\alpha))
  = \lim_{z\rightarrow \zeta(\alpha)^-}
  \re\left\{i\zeta'(\alpha)[u(z)-iv(z)]\right\} \\
  &= \frac{1}{2}H\gamma(\alpha) + \frac{1}{2\pi}\int_0^{2\pi}
  [G_1(\alpha,\beta) + G_2(\alpha,\beta)]\gamma(\beta)\,d\beta.
\end{aligned}
\end{equation}
Here $G_j(\alpha,\beta) = \re\{\wtil G_j(\alpha,\beta)\}$ and
$i\zeta'(\alpha)/|\zeta'(\alpha)|$ represents the normal vector to the
curve. Note that the dot product of two complex numbers $z$ and $w$
(thought of as vectors in $\mathbb{R}^2$) is $\re\{z\bar w\}$.
Once $\mc G\varphi(x)$ is known, we can evaluate the right hand
side of (\ref{eq:ww}) using (\ref{eq:uv:from:G}).

For visualization, it is often useful to evaluate the velocity and
pressure inside the fluid.  The velocity was already given in
terms of the vortex sheet strength in (\ref{eq:dPhi1}) above.
For pressure, we use the unsteady Bernoulli equation
\begin{equation}\label{eq:pressure}
  \phi_t + \frac{1}{2}|\nabla\phi|^2 + gy + \frac{p}{\rho} = c(t),
\end{equation}
where $c(t)$ was given in (\ref{eq:c:def}).  One option for computing
$\phi_t$ is to differentiate (\ref{eq:fred2}) with respect to time to
obtain an integral equation for $\mu_t$ (see \cite{water:pod}), then
express $\phi_t$ in terms of $\mu_t$ by differentiating
(\ref{eq:Phi}). A simpler approach is to differentiate the Laplace
equation (\ref{eq:dno}) with respect to time.  The value of $\phi_t$
on the upper boundary is $\varphi_t - \phi_y\eta_t$, which is known.
Since the real part of (\ref{eq:Phi}) gives the solution $\phi(z)$ of
Laplace's equation with boundary condition $\varphi$ on the upper
surface, we can replace $\varphi$ with $\varphi_t - \phi_y\eta_t$ in
(\ref{eq:fred2}) to convert (\ref{eq:Phi}) into a formula for
$\phi_t(z)$ instead.

\section{Linearized and adjoint equations for the water wave}
\label{sec:adjoint}

In this section we derive explicit formulas for the variational and
adjoint equations of Sections~\ref{sec:acm} and~\ref{sec:trust}.  A
dot will be used to denote a directional derivative with respect to
the initial conditions.  The equation $\dot q_t = DF(q)\dot q$ of
(\ref{eq:dot}) is simply
\begin{subequations}\label{eq:lin2}
  \begin{align}
    \dot\eta(x,0)=\dot\eta_0(x),\quad \dot\varphi(x,0) =\dot\varphi_0(x),&&
    t& =0,\label{eq:l1}\\
    \dot\phi_{xx}+\dot\phi_{yy} =0,&&
    -h& <y<\eta,\label{eq:l2}\\
    \dot\phi_y =0,&&
    y& =-h,\label{eq:l3}\\
    \dot\phi+\phi_y\dot\eta =\dot\varphi,&&
    y& =\eta,\label{eq:l4}\\
    \dot\eta_t+\dot\eta_x\phi_x+\eta_x\dot\phi_x+
    \eta_x\phi_{xy}\dot\eta =\dot\phi_y+\phi_{yy}\dot\eta,&&
    y& =\eta,\label{eq:l5}\\
    \dot\varphi_t
    = P\left[-\bigg(\eta_x\phi_x\phi_y + \frac{1}{2}\phi_x^2 -
        \frac{1}{2}\phi_y^2\bigg)^\text{\large .} - g\dot\eta +
      \frac{\sigma}{\rho}\partial_x\left(\frac{\dot\eta_x}{(1+\eta_x^2)^{3/2}}\right)\right],&&
    y&=\eta.\label{eq:l6}
  \end{align}
\end{subequations}
Note that evaluation of $\dot\phi(x,y,t)$ on the free surface gives
$\big[\dot\varphi(x)-\phi_y(x,\eta(x),t)\dot\eta(x)\big]$ rather than
$\dot\varphi(x)$ due to the boundary perturbation.  Making use of
$\phi_{yy}=-\phi_{xx}$, (\ref{eq:l5}) can be simplified to
\begin{equation}\label{eq:l5a}
  \dot\eta_t = \big(\dot\phi_y - \eta_x\dot\phi_x\big) -
  \big(\dot\eta\phi_x\big)',
\end{equation}
where a prime indicates an $x$-derivative along the free surface,
e.g.~$f':=\frac{d}{dx}f(x,\eta(x),t)=f_x+\eta_x f_y$.  Equation
(\ref{eq:l6}) may also be simplified, using
\begin{align}
\notag
  \bigg(\eta_x\phi_x\phi_y + \frac{1}{2}\phi_x^2-\frac{1}{2}\phi_y^2
      \bigg)^\text{\large .} &=
      \dot\eta_x\phi_x\phi_y + \eta_x\dot\phi_x\phi_y +
      \eta_x\phi_{xy}\dot\eta\phi_y + \eta_x\phi_x\dot\phi_y +
      \eta_x\phi_x\phi_{yy}\dot\eta + \phi_x\dot\phi_x + \phi_x\phi_{xy}\dot\eta
      -\phi_y\dot\phi_y - \phi_y\phi_{yy}\dot\eta \\
      \label{eq:l6a}
      &= \big(\dot\eta\phi_x\phi_y\big)' + \phi_x\dot\phi' -
      \phi_y\big(\dot\phi_y - \eta_x\dot\phi_x\big).
\end{align}
The equation $\tilde{q}_s=D F(q)^*\tilde{q}$ is obtained from
\begin{equation}\label{eq:tilde1}
  \begin{split}
    \langle\dot{q},\tilde{q}_s\rangle=\langle\dot{q}_t,\tilde{q}\rangle&=
    \frac{1}{2\pi}\int_0^{2\pi}\left[\underline{(\dot\phi_y-\eta_x\dot\phi_x)}
      - (\dot\eta\phi_x)'\right]
    \tilde\eta\,dx\\
    &\qquad+\frac{1}{2\pi}\int_0^{2\pi}P\left[-(\dot\eta\phi_x\phi_y)'-
      \phi_x\dot\phi'+\phi_y\underline{(\dot\phi_y-\eta_x\dot\phi_x)}
      -g\dot\eta+\frac{\sigma}{\rho}\partial_x\left(
        \frac{\dot\eta_x}{(1+\eta_x^2)^{3/2}}\right)\right]\tilde\varphi\,dx.
\end{split}
\end{equation}
The right-hand side must now be re-organized so we can identify
$\tilde q_s$.  $P$ is self-adjoint, so it can be transferred from the
bracketed term to $\tilde\varphi$.  The underlined terms may be
written $\mc G\dot\phi$, where $\dot\phi$ is evaluated on the free
surface.  Green's identity shows that $\mc G$ is self-adjoint. Indeed,
if $\dot\phi$ and $\chi$ satisfy Laplace's equation with Neumann
conditions on the bottom boundary, then
\begin{equation}
  0 = \iint(\chi\Delta\dot\phi-\dot\phi\Delta\chi)d A =
  \int\chi\frac{\partial\dot\phi}{\partial n}-
  \dot\phi\frac{\partial\chi}{\partial n} ds = 
  \int\chi\mc G\dot\phi\,dx - \int\dot\phi\mc G\chi\,dx.
\end{equation}
Thus, from (\ref{eq:tilde1}), we obtain
\begin{equation*}
    \langle\dot{q},\tilde{q}_s\rangle = \frac{1}{2\pi}\int_0^{2\pi}
    \left[ \dot\phi\mc G\chi +
      \dot\eta\phi_x\tilde\eta' + \dot\eta\phi_x\phi_y (P\tilde\varphi)'
      + \dot\phi(\phi_x P\tilde\varphi)' - g\dot\eta P\tilde\varphi
      + \frac{\sigma}{\rho}\dot\eta \partial_x\left( \frac{
          \tilde\varphi_x}{(1+\eta_x^2)^{3/2}} \right)\right]dx,
\end{equation*}
where $\chi$ is an auxiliary solution of Laplace's equation defined to be
$\big(\tilde\eta + \phi_y P\tilde\varphi\big)$ on the free surface.
Finally, we substitute $\dot\phi = \dot\varphi - \phi_y\dot\eta$ and
match terms to arrive at the adjoint system
\begin{subequations}\label{eq:adj2}
  \begin{align}
    \tilde\eta(x,0)=0,\quad
    \tilde\varphi(x,0) =\varphi(x,T/4),&&
    s&=0,\label{eq:a1}\\
    \chi_{xx}+\chi_{yy} =0,&&
    -h&<y<\eta,\label{eq:a2}\\
    \chi_y =0,&&
    y&=-h,\label{eq:a3}\\
    \chi =\tilde\eta+\phi_y P \tilde\varphi,&&
    y&=\eta,\label{eq:a4}\\
    \tilde\varphi_s= (\chi_y - \eta_x\chi_x) +
    (\phi_x P\tilde\varphi)',&&
    y&=\eta,\label{eq:a5}\\
    \tilde\eta_s = -\phi_y(\chi_y-\eta_x\chi_x)
    + \phi_x\tilde\eta_x - \phi_y\phi_x'P\tilde\varphi
    - g P\tilde\varphi+\frac{\sigma}{\rho}\partial_x
    \left(\frac{\tilde\varphi_x}{(1+\eta_x^2)^{3/2}}\right),&&
    y&=\eta.\label{eq:a6}
\end{align}
\end{subequations}
The initial conditions (\ref{eq:a1}) are specific to the objective
function (\ref{eq:f:phi:again}), but are easily modified to handle
the alternative objective function
(\ref{eq:f:naive}).  Note that the adjoint problem has the same
structure as the forward and linearized problems, with a Dirichlet to
Neumann map appearing in the evolution equations for $\tilde\eta$ and
$\tilde\varphi$. We use the boundary integral method described in
Appendix~\ref{sec:BI} to compute $\mc G\chi$, and employ a dense
output formula to interpolate $\eta$ and $\varphi$ between timesteps
at intermediate Runge-Kutta stages of the adjoint problem, as
explained in Section~\ref{sec:acm}.

\section{Levenberg-Marquardt implementation with delayed Jacobian
updates}
\label{sec:levmar}

Since minimizing $f$ in (\ref{eq:f:phi}) is a small-residual nonlinear
least squares problem, the Levenberg-Marquardt method \cite{nocedal}
is quadratically convergent. Our goal in this section is to discuss
modifications of the algorithm in which re-computation of the Jacobian
is delayed until the previously computed Jacobian ceases to be useful.
By appropriately adjusting the step size in the numerical continuation
algorithm, it is usually only necessary to compute the Jacobian once
per solution. Briefly, the Levenberg-Marquardt method works by
minimizing the quadratic function
\begin{equation}
  f_\text{approx}(p) = f(c) + g^Tp + \frac{1}{2}p^TBp,
  \qquad g = \nabla f(c) = J^T(c)r(c), \qquad B = J(c)^TJ(c)
\end{equation}
over the trust region $\|p\|\le\Delta$. The true Hessian of $f$ at $c$
satisfies $H-B=\sum_i r_i\nabla^2r_i$, which is small if $r$ is small.
The solution of this constrained quadratic minimization problem is the
same as the solution of a linear least-squares problem with an unknown
parameter $\lambda$:
\begin{equation}
  \min_p\left\|\begin{pmatrix} J \\ \sqrt{\lambda}\,I \end{pmatrix} p
      + \begin{pmatrix} r \\ 0 \end{pmatrix} \right\|, \qquad
    \lambda\ge 0, \qquad (\|p\|-\Delta)\lambda=0.
\end{equation}
Formulating the problem this way (instead of solving $(B+\lambda
I)p=-g$) avoids squaring the condition number of $J$.  Rather than use
the MINPACK algorithm \cite{nocedal} to find the Lagrange multiplier
$\lambda$, we compute the (thin) SVD of $J$, and define
\begin{equation}
  J = USV^T, \qquad S=\opn{diag}\{\sigma\}, \qquad
  \tilde p=V^Tp, \qquad \tilde r = U^T r, \qquad
  \tilde g = S^T\tilde r.
\end{equation}
Here $U$ is $m\times n$ and $S=S^T$ is $n\times n$.  This leads to an
equivalent problem
\begin{equation}\label{eq:trust:tilde}
  \min_{\tilde p}\left\|\begin{pmatrix} S \\ \sqrt{\lambda}\,I \end{pmatrix}
    \tilde p
    + \begin{pmatrix} \tilde r \\ 0 \end{pmatrix} \right\|, \qquad
  \lambda\ge 0, \qquad (\|\tilde p\|-\Delta)\lambda=0,
\end{equation}
which can be solved in $O(n)$ time by performing a Newton iteration on
$\tau(\lambda)$, defined as
\begin{equation}
  \tau(\lambda)=\frac{1}{\|\tilde p\|}-\frac{1}{\Delta}, \qquad
  \tilde p = \arg\min\left\|\begin{pmatrix} S \\
      \sqrt\lambda\,I \end{pmatrix} \tilde p +
    \begin{pmatrix} \tilde r \\ 0 \end{pmatrix} \right\|.
\end{equation}
It is easy to show that $\tau$ is an increasing, concave down function
for $\lambda\ge0$ (assuming $S$ is non-singular); thus, if
$\tau(0)<0$, the Newton iteration starting at $\lambda^\e 0=0$ will
increase monotonically to the solution of (\ref{eq:trust:tilde}) with
$\tau(\lambda^\e l)$ increasing to zero.  This Newton iteration is
equivalent to
\begin{center}
\fbox{
\parbox{4in}{
\begin{tabbing}
\hspace*{.25in} \= \hspace*{.25in} \= \kill
$l=0$,\quad $\lambda^\e 0=0$,\quad $\tilde p_0 = \arg\min_{\tilde p}
\left\|S\tilde p + \tilde r\right\|$ \\
\textbf{while} $\jd\left(\frac{\|\tilde p_l\|-\Delta}{\Delta}\right) >
\text{tol}$ \\
\> $\jd\lambda^\e{l+1} = \lambda^\e l + \frac{\tilde p_l^T\tilde p_l}{
  \tilde p_l^T(S^TS+\lambda^\e l I)\tilde p_l}\left(
  \frac{\|\tilde p_l\| - \Delta}{\Delta}\right)$ \\
\> $l=l+1$ \\
\> $\tilde p_l = \arg\min_{\tilde p}
  \left\|\begin{pmatrix} S \\ \sqrt{\lambda^\e l}\,I \end{pmatrix} \tilde p +
    \begin{pmatrix} \tilde r \\ 0 \end{pmatrix} \right\|$ \\
\textbf{end}
\end{tabbing}
}}
\end{center}
We use $\text{tol}=10^{-12}$ in double-precision and $10^{-24}$ in
quadruple precision.  It is not critical that $\lambda$ be computed to
such high accuracy, but as the Newton iteration is inexpensive
once the SVD of $J$ is known, there is no reason not to iterate to
convergence. At the end, we set $p=V\tilde p$.

We remark that it is more common to compute $\lambda$ by a sequence of
QR factorizations of $[J;\sqrt{\lambda^\e l}\,I]$, as is done in
MINPACK.  However, the SVD approach is simpler, and similar in
speed, since several QR factorizations have to be performed to compute
$\lambda$ while only one SVD must be computed.  Moreover, we can
re-use $J$ several times instead of re-computing it each time a step
is accepted.  When this is done, it pays to have factored $J=USV^T$ up
front.

Delaying the computation of $J$ requires a modified strategy for
updating the trust region radius, as well as a means of deciding when
the minimization is complete, and when to re-compute $J$.  Our
design decisions are summarized as follows:

1. The algorithm terminates if $f=0$, or if $c$ is unchanged from the
previous iteration (i.e.~$c+p$ equals $c$ in floating point
arithmetic), or if the algorithm reaches the \emph{roundoff\_regime}
phase, and then a step is rejected or \emph{stepsJ} reaches
\emph{max\_stepsJ}.  Here \emph{stepsJ} counts accepted steps since
$J$ was last evaluated, and the \emph{roundoff\_regime} phase begins
if $f<f_\text{tol}$ or $\Delta<g_\text{tol}$, where the tolerances and
\emph{max\_stepsJ} are specified by the user.  If the Jacobian has just been
computed (i.e.~$\text{\emph{stepsJ}}=0$), we also check if
$\|g\|<g_\text{tol}$ or $|df|/f<df_\text{tol}$ to trigger
\emph{roundoff\_regime}. Here $df=f_\text{approx}(c+p)-f(c)$ is the
predicted change in $f$ when minimizing the quadratic model
$f_\text{approx}$ over the trust region, and $df_\text{tol}$ is
specified by the user. We used
\begin{equation*}
f_\text{tol}=10^{-26}, \qquad g_\text{tol}=10^{-13}, \qquad
\text{\emph{max\_stepsJ}}=10, \qquad df_\text{tol} = 10^{-5}.
\end{equation*}
The idea of \emph{roundoff\_regime} is to try to improve $f$ through a
few additional residual calculations without recomputing~$J$.

2. Steps are accepted if $\rho=[f(c+p)-f(c)]/df>0$; otherwise they are
rejected.  Note that $\rho$ is the ratio of the actual change to the
predicted change, the latter being negative.  We also use $\rho$ to
adjust $\Delta$.  If $\rho<\rho_0=1/4$, we replace $\Delta$ by $\|p\|$
times $\alpha_0=3/8$.  If $\rho>\rho_1=0.85$ and $\|p\|>0.9\Delta$, we
multiply $\Delta$ by $\alpha_1=1.875$.  Otherwise we leave $\Delta$
alone. So far this agrees with the standard trust region mechanism
\cite{nocedal} for adjusting $\Delta$, with slightly different
parameters.  What we do differently is define a parameter
\emph{delta\_trigger} to be a prescribed fraction, namely
$\alpha_2=0.2$, of \emph{delta\_first\_rejected}, the first rejected
radius after (or coinciding with) an accepted step. Note that the
radius is rejected ($\rho<\rho_0$), not necessarily the step
($\rho\le0$).  The reason to wait for an accepted step is to let the
trust region shrink normally several times in a row if the Jacobian is
freshly computed ($\text{\emph{stepsJ}}=0$).

3. The Jacobian is re-computed if
\emph{roundoff\_regime} has not occurred, and either \emph{stepsJ}
reaches \emph{max\_stepsJ}, or $\text{\emph{stepsJ}}>0$ and $\Delta$
drops below \emph{delta\_trigger}, or $\text{\emph{stepsJ}}>0$ and
$|df|/f<df_\text{tol}$.  This last test avoids iterating on an old
Jacobian if the new residual is nearly orthogonal to its columns ---
there is little point in continuing if $f_\text{approx}$ cannot be
decreased significantly.  The parameters $\alpha_i$ were chosen so
that
\begin{equation}
  \max(\alpha_0^2,\alpha_0^3\alpha_1^2)<\alpha_2<
  \min(\alpha_0,\alpha_0^2\alpha_1),
\end{equation}
which triggers the re-computation of $J$ if two radii are rejected in
a row, or on a reject-accept-reject-accept-reject sequence, assuming
$\|p\|=\Delta$ on each rejection.  Before computing $J$, if
\emph{delta\_first\_rejected} has been defined since $J$ was last
computed, we reset $\Delta$ to
\begin{equation*}
  \Delta = \text{\emph{delta\_first\_rejected}}/\alpha_1.
\end{equation*}
This makes up for the decreases in $\Delta$ that occur due to using an
old Jacobian.

4. We compute $r$ but not $J$ if a step is rejected on a freshly
computed Jacobian, or if a step is accepted or rejected without
triggering one of the conditions mentioned above for computing $J$.

\vspace*{.25in}
\noindent\textbf{References}
%% References


\bibliographystyle{elsarticle-num}

%\bibliography{refs.bib}

\end{document}